\documentclass{emulateapj}

\newcommand{\Spitzer}{{\sl Spitzer}}

\newcommand{\HST}{{\sl HST}}

\newcommand{\Msun}{\mbox{M$_{\sun}$}}

\newcommand{\Lsun}{\mbox{L$_{\sun}$}}
\newcommand{\Rsun}{\mbox{R$_{\sun}$}}
\newcommand{\Mjup}{\mbox{M$_{\rm Jup}$}}

\newcommand{\etal}{et al.}

\newcommand{\kms}{\hbox{km~s$^{-1}$}}

\newcommand{\Kp}{\mbox{$K^{\prime}$}}
\newcommand{\Ks}{\mbox{$K_S$}}

\newcommand{\degs}{\mbox{$^{\circ}$}}
\newcommand{\Mtot}{\mbox{$M_{\rm tot}$}}
\newcommand{\Lbol}{\mbox{$L_{\rm bol}$}}

\newcommand{\Teff}{\mbox{$T_{\rm eff}$}}
\newcommand{\logg}{\mbox{$\log(g)$}}

\newcommand{\Lp}{\mbox{${L^\prime}$}}

\newcommand{\twomassbin}{\hbox{2MASS~J1534$-$2952AB}}
\newcommand{\hdbin}{\hbox{HD~130948BC}}

\newcommand{\lhsint}{\hbox{LHS~2397a}}
\newcommand{\lhsbin}{\hbox{LHS~2397aAB}}
\newcommand{\lhsA}{\hbox{LHS~2397aA}}
\newcommand{\lhsB}{\hbox{LHS~2397aB}}
\newcommand{\orbit}{\hbox{\tt ORBIT}}

\slugcomment{Submitted to ApJ 2009 February 9; accepted 2009 April 14}

\shorttitle{Dynamical Mass of LHS 2397aAB}
\shortauthors{Dupuy \etal}

\begin{document}

\title{Keck Laser Guide Star Adaptive Optics Monitoring of the M8+L7
  Binary LHS~2397\lowercase{a}AB: First Dynamical Mass Benchmark at
  the L/T Transition \altaffilmark{*,\dag,\ddag,\S}}

\author{Trent J.\ Dupuy, Michael C.\ Liu\altaffilmark{1}}
\affil{Institute for Astronomy, University of Hawai`i, 2680 Woodlawn
  Drive, Honolulu, HI 96822} 
\and
\author{Michael J.\ Ireland}
\affil{School of Physics, University of Sydney, NSW 2006, Australia}

      \altaffiltext{*}{Some of the data presented herein were obtained
        at the W.M. Keck Observatory, which is operated as a
        scientific partnership among the California Institute of
        Technology, the University of California, and the National
        Aeronautics and Space Administration. The Observatory was made
        possible by the generous financial support of the W.M. Keck
        Foundation.}

      \altaffiltext{\dag}{Based partly on observations made with the
        NASA/ESA {\sl Hubble Space Telescope}, obtained from the data
        archive at the Space Telescope Institute. STScI is operated by
        the association of Universities for Research in Astronomy,
        Inc. under the NASA contract NAS 5-26555.}

      \altaffiltext{\ddag}{Based partly on observations obtained under
        program ID GN-2002A-Q-6 at the Gemini Observatory, which is
        operated by the Association of Universities for Research in
        Astronomy, Inc., under a cooperative agreement with the NSF on
        behalf of the Gemini partnership: the National Science
        Foundation (United States), the Science and Technology
        Facilities Council (United Kingdom), the National Research
        Council (Canada), CONICYT (Chile), the Australian Research
        Council (Australia), Minist\'{e}rio da Ci\^{e}ncia e
        Tecnologia (Brazil) and SECYT (Argentina).}
      
      \altaffiltext{\S}{Based partly on observations made with ESO
        Telescopes at the Paranal Observatory under program IDs
        071.C-0716 and 076.C-0139.}

      \altaffiltext{1}{Alfred P. Sloan Research Fellow}

\begin{abstract}

  We present Keck laser guide star adaptive optics imaging and
  aperture masking observations of the M8+L7 binary \lhsbin.  Together
  with archival \HST, Gemini-North, and VLT data, our observations
  span 11.8~years of the binary's 14.2-year orbital period.  We
  determine a total dynamical mass of 0.146$^{+0.015}_{-0.013}$~\Msun\
  (153$^{+16}_{-14}$~\Mjup).  Using the combined observational
  constraints of the total mass and individual luminosities, the
  Tucson (Lyon) evolutionary models give an age for the system of
  1.5$^{+4.1}_{-0.6}$~Gyr (1.8$^{+8.2}_{-0.8}$~Gyr), which is
  consistent with its space motion based on a comparison to the
  Besan\c{c}on Galactic structure model.  We also use these models to
  determine the mass ratio, giving individual masses of
  0.0839$^{+0.0007}_{-0.0015}$~\Msun\
  (0.0848$^{+0.0010}_{-0.0012}$~\Msun) for \lhsA\ and
  0.061$^{+0.014}_{-0.011}$~\Msun\ (0.060$^{+0.008}_{-0.012}$~\Msun)
  for \lhsB. Because \lhsB\ is very close to the theoretical
  mass-limit of lithium burning, which remains untested by dynamical
  masses, measuring its lithium depletion would uniquely test
  substellar models.  We estimate a spectral type of L7$\pm$1 for
  \lhsB, making it the first L/T transition object with a dynamical
  mass determination.  This enables a precise estimate of its
  effective temperature from Tucson (Lyon) evolutionary models of
  1450$\pm$40~K (1430$\pm$40~K), which is 200~K higher than estimates
  for young late-L companions but consistent with older late-L field
  dwarfs, supporting the idea that the temperature of the L/T
  transition is surface gravity dependent. Comparing our temperature
  estimate for \lhsB\ to those derived from spectral synthesis
  modeling for similar objects reveals consistency between
  evolutionary and atmospheric models at the L/T transition, despite
  the currently limited understanding of this phase of substellar
  evolution.  Future dynamical masses for L/T binaries spanning a
  range of surface gravity, age, and mass will provide the next
  critical tests of substellar models at the L/T transition.

\end{abstract}



\section{Introduction}

In the field population, the warmest brown dwarfs and the lowest mass
stars are in many ways quite different. Beyond the fact that brown
dwarfs by definition do not sustain internal energy generation through
hydrogen fusion, while very low-mass stars do, the effective
temperatures of these two classes of objects span a range of
$\approx$1500~K and more than an order of magnitude in luminosity.
However, they share one important attribute, which is the presence of
dust clouds in their atmospheres. Below effective temperatures of
$\approx$2800~K, photospheres are cool enough for particulate matter
to form \citep[e.g.,][]{1996A&A...308L..29T, 1997ApJ...480L..39J}, and
this dust persists for objects as cool as $\approx$1300~K
\citep[e.g.,][]{2000ApJ...542..464C}, at which temperature the dust
begins to settle below the photosphere. Thus, the formation and
sedimentation of dust clouds are among the key processes driving the
atmospheric physics of objects in this temperature range, which also
includes the warmest extrasolar planets such as transiting hot
Jupiters and young giant planets \citep[e.g.,][]{2008ApJ...676L..61B,
  2008Sci...322.1348M}.

In addition to providing excellent laboratories for ultracool
atmospheric physics, very low-mass stars and brown dwarfs are also
intrinsically interesting as the lowest mass products of star
formation. In particular, the properties of binary systems spanning a
wide range of stellar masses can provide discriminating tests of
various star formation models. While the binary properties of low-mass
stars (FGKM spectral types) have been extensively studied
\citep[e.g.,][]{1991A&A...248..485D, 1992ApJ...396..178F}, such
studies have only become possible within the past decade for brown
dwarfs and stars at the bottom of the main sequence. In fact, such
studies are generally hampered by the fact that brown dwarfs follow a
mass--luminosity--age relation, rather than the simpler
mass--luminosity relation for main-sequence stars. For example, a
late-M field dwarf may be a young brown dwarf or a star at the bottom
of the main sequence. Binaries with well determined orbits are
extremely valuable as they break this mass--age degeneracy by
providing dynamical mass estimates, though very few mass
determinations are available to date
\citep[e.g.,][]{2004astro.ph..7334O, 2009ApJ...692..729D}. In fact,
mass measurements from binaries typically offer much stronger
constraints on substellar models than other commonly measured
parameters (e.g., age and compositon) given that their higher
precision, hence the concept of using binaries as ``mass benchmarks''
\citep{2008ApJ...689..436L}. In addition, binary orbits provide a
wealth of other information such as the true (not projected) semimajor
axis and the eccentricity, both of which have different predicted
distributions under different formation scenarios
\citep[e.g.,][]{2009MNRAS.392..590B, 2004ApJ...600..769F}.

One particularly striking result of the many surveys that have
searched for brown dwarf binaries is the paucity of intermediate mass
ratio binaries. Although such surveys have typically been sensitive to
mass ratios as small as $M_{\rm B}/M_{\rm A}\approx$~0.6, the mass
ratio distribution drops sharply below unity, with $>$~50\% of
binaries having estimated mass ratios of $>$~0.9
\citep{2007prpl.conf..427B}. This is in stark contrast to solar-type
binaries whose mass ratio distribution peaks at $\approx$0.4
\citep{1991A&A...248..485D}, suggesting a different formation process
for these two classes of objects. Although very unequal mass (or
spectral type) brown dwarf binaries are rare, they provide the
strongest coevality tests of substellar theoretical models because
they offer the most leverage on constraining model isochrones.

\citet{2003ApJ...584..453F} discovered a faint companion to
\lhsint\footnote[2]{\lhsint\ is the star listed between LHS~2397 and
  LHS~2398 in the LHS Catalog \citep{1979lccs.book.....L}.} using the
curvature AO system Hokupa`a at Gemini-North, revealing one of the
most unequal-flux ultracool binaries known
($\Delta{K}$~=~2.77$\pm$0.10~mag). \lhsint\ (M8) is an H$\alpha$ flare
star with a distance measured by \citet{1992AJ....103..638M} to be
14.3$\pm$0.4~pc. \citet{2003ApJ...584..453F} estimated the spectral
type of \lhsB\ from near-infrared photometry to be L7.5$\pm$1. Thus,
\lhsbin\ represents a rare pairing of two objects at the extreme ends
of the effective temperature range over which dust clouds play an
important role in atmospheric physics. \lhsA\ is just cool enough that
dust is likely to persist in its atmosphere, while \lhsB\ has nearly
cooled to the point of becoming a dust-free T~dwarf.

Based on Keck laser guide star adaptive optics (LGS AO) imaging and
aperture masking observations from our ongoing orbital monitoring
program of ultracool binaries, we present here a dynamical mass for
\lhsbin. Combining our Keck data with archival {\sl Hubble Space
  Telescope} (\HST), Very Large Telescope (VLT), and Gemini-North
Telescope images, we measure a total mass of 0.146$\pm$0.014~\Msun,
with the dominant source of uncertainty being the 3.0\% error in the
parallax (which translates into a 9.0\% error in the mass). This is
the first mass measurement for a binary containing an L/T transition
brown dwarf, and we use the direct mass measurement along with the
measured luminosities of both components to test evolutionary and
atmospheric models in this temperature regime ($\approx$1400~K) for
the first time.


\section{Observations \label{sec:obs}}

\subsection{\HST/WFPC2-PC1  \label{sec:hst}}

We retrieved \HST\ archival images of \lhsbin\ obtained with the WFPC2
Planetary Camera (PC1) on UT 1997 April 12 (GO-6345, PI Kirkpatrick).
The data comprise four F814W images with exposure times of 2~s,
1000~s, 1000~s, and 300~s, taken in that order in a single orbit. The
companion \lhsB\ is only seen in the latter three images, and the
primary star \lhsA\ is saturated in all but the first image. We used
TinyTim \citep{1995ASPC...77..349K} to generate point-spread-function
(PSF) models which were fit to the data in a similar fashion to our
previous work \citep{2008ApJ...689..436L, 2009ApJ...692..729D}, thus
allowing us to determine the positions and fluxes of \lhsbin\ in each
image. Saturated pixels were masked in our PSF-fitting routine. The
F814W flux ratio of the binary (4.18$\pm$0.08~mag) was determined by
comparing the mean peak flux (i.e., the model PSF's normalization
constant) of \lhsB\ in the latter three images to the peak flux of
\lhsA\ in the first image.

Since positional information is carried almost entirely in the core of
the PSF, especially for these undersampled WFPC2 images, we could not
determine precise positions for \lhsA\ in the three images in which
its core was saturated. Therefore, in order to measure the binary
separaton and P.A., we had to somehow tie the latter three
measurements of \lhsB\ to the first shallow image of \lhsA. This was
accomplished by fitting a first-order polynomial to the positions of
\lhsB, in order to account for any telescope drift that occurred
during the observations, and extrapolating this fit to the unsaturated
image of \lhsA. We found a drift in the $(x, y)$ position of \lhsB\ of
(0.82$\pm$0.03, 0.26$\pm$0.01)~pix/hr over the 0.7~hr duration of the
observations.  After correcting for this drift and using a pixel scale
of 45.54$\pm$0.01~mas/pix,\footnote[3]{This is the quoted pixel scale
  from the WFPC2 Instrument Handbook for Cycle 13, which is consistent
  with other values in the literature \citep[e.g., see the discussion
  by][]{2008ApJ...689..436L}.} we found a separation of 274$\pm$4~mas,
and a position angle (P.A.) of 87.3$\pm$0.8\degs\
(Table~\ref{tbl:astrom}).

\begin{figure*}
\plotone{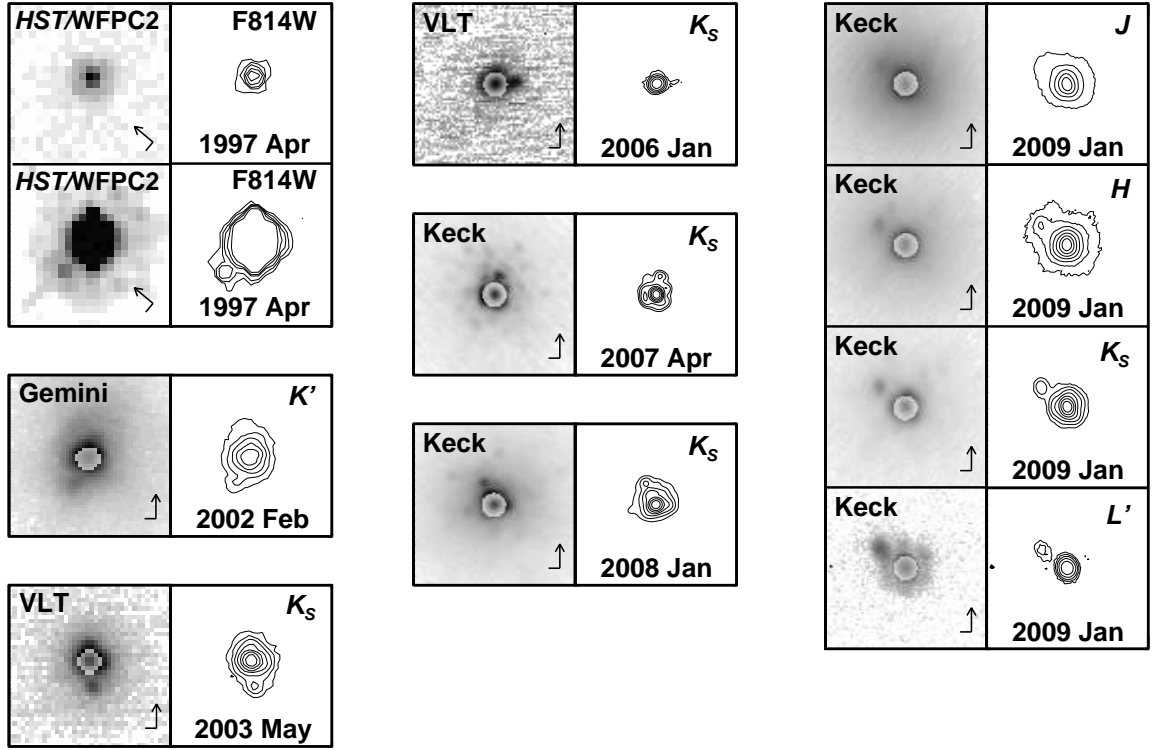}
\caption{ \normalsize \HST, Gemini, VLT and Keck images of \lhsbin\
  shown chronologically by column.  We do not rotate the \HST\ data so
  that north is up in order to preserve the somewhat undersampled
  nature of the WFPC2 data.  The Airy ring of the Keck PSF is visible
  in some Keck images.  All images are shown on the same scale,
  1$\farcs$0 on a side, using a square-root stretch for the grayscale
  images.  Pixels within a 75~mas radius of the primary have been
  divided by 10 to more clearly show the companion.  Contours are
  drawn at 0.62, 0.38, 0.24, 0.15, 0.090, 0.055, and 0.034 of the peak
  pixel.  The two lowest contours are not drawn for the \HST, Gemini,
  VLT (2006), and Keck $J$- and \Lp-band images . \label{fig:data}}
\end{figure*}

To determine the uncertainties of our position and flux measurements
we simulated the data using images of single stars, allowing us to
assess systematic errors in our fitting procedure. We built a library
of single stars from other programs that targeted brown dwarfs with
\HST/WFPC2-PC1 (GO-8563, PI Kirkpatrick; GO-8581, PI Reid; and
GO-8146, PI Reid). We only used stars that had equivalent or higher
signal-to-noise (S/N) than the science data so that we could degrade
the S/N of the library images to match the science data. We also
restricted ourselves to observations consisting of two or more images
to allow robust rejection of cosmic rays. This library served to
simulate the unsaturated images of \lhsA\ and \lhsB. To simulate the
saturated images of \lhsA\ we used other targets from the same program
that had also been intentionally saturated to search for faint
companions. We used the five targets closest in apparent flux to
\lhsA\ (LHS~2924, BRI~0021-0214, TVLM~513-46546, LP~412-31, and
LHS~2243), which were $\lesssim$~30\% different in flux as measured by
our PSF-fitting routine. We did not scale these images to match
\lhsA's flux since it is impossible to scale the saturated pixels. We
added images of single stars from our library to these saturated
images after scaling them down and adding Poisson noise to match the
S/N of \lhsB. We shifted these images by an integer number of pixels
that best reproduced the observed separation of ($+$4.1, $+$4.4)
pixels. By applying our PSF-fitting routine to these images, we were
able to assess the position and flux measurements of a faint companion
made in the presence of a saturated primary.

From these Monte Carlo simulations, we derived root-mean-square (rms)
errors in measured $(x, y)$ positions of (0.05, 0.06)~pix and (0.07,
0.07)~pix for the 300~s and 1000~s images, respectively, with no
significant systematic offsets.\footnote[4]{We found significant
  statistical correlation between the truth-minus-fitted offsets of
  the $x$ and $y$ positions
  ($\sigma_{xy}^2/\sigma_x\sigma_y\approx0.6$). We accounted for this
  covariance when propagating errors to compute the uncertainty in the
  separation and P.A.} For the fluxes, the ratio of the input to
best-fit value was 1.05$\pm$0.05 for the 300~s image and 1.07$\pm$0.11
for the 1000~s images. In addition, we were able to assess the
intrinsic uncertainty in our TinyTim PSF-fitting routine (i.e., the
error when \emph{not} in the presence of a saturated star) by
comparing each undithered pair of PSF library images taken in the same
orbit. From this ensemble of single-star position and flux
measurements we found that the scatter in both $x$ and $y$ position
measurements was 0.03~pix, and the scatter in peak flux was a factor
of 0.06. We adopt these errors for the position and flux of \lhsA\
measured in the 2~s image, and we adopt the errors from the Monte
Carlo for the positions and fluxes of \lhsB\ measured in the longer,
saturated-primary exposures. This results in a separation uncertainty
of 0.08~pix (4~mas), a P.A. uncertainty of 0.8\degs, and a flux ratio
uncertainty of 0.08~mag.

The drift correction we derived is thus quite significant compared to
our derived positional uncertainties. However, if we do not apply this
drift correction, the scatter in the measured position of \lhsB\ is
twice as large as predicted from our Monte Carlo simulations.  As
discussed in Section~\ref{sec:orbit}, the orbit determination results
in an almost identical dynamical mass if this single \HST\ epoch is
excluded. Thus, our derived relative astrometry for the \HST\ data is
consistent with the orbit as determined from all other available
astrometry.

\subsection{Keck/NIRC2 LGS \label{sec:keck}}

We have monitored \lhsbin\ using the laser guide star adaptive optics
(LGS AO) system at the Keck~II Telescope on Mauna Kea, Hawaii
\citep{2006PASP..118..297W, 2006PASP..118..310V}. We used the facility
near-infrared camera NIRC2 in its narrow field-of-view mode because
this provides the finest pixel scale. At each epoch, we obtained data
in one or more filters covering the standard atmospheric windows from
the Mauna Kea Observatories (MKO) filter consortium
\citep{mkofilters1, mkofilters2}. We also obtained data with the
$CH_4s$ filter, positioned shortward of a methane absorption band
($\lambda_c$~=~1.592~\micron, $\Delta\lambda$~=~0.126~\micron), to
enable more robust photometric discrimination in the spectral type
estimate of \lhsB.

The LGS provided the wavefront reference source for AO correction,
with the exception of tip-tilt motion.  The LGS brightness, as
measured by the flux incident on the AO wavefront sensor, was
equivalent to a $V$~$\approx$~10.4--10.7~mag star.  The tip-tilt
correction and quasi-static changes in the image of the LGS as seen by
the wavefront sensor were measured contemporaneously by a second,
lower bandwidth wavefront sensor monitoring \lhsint, which saw the
equivalent of a $R$~$\approx$~15.7--16.2~mag star.\footnote[5]{The
  Keck LGS AO tip-tilt sensor is very red-sensitive, so a red source
  like \lhsint\ appears much brighter than would otherwise be expected
  \citep[$R$~=~17.9~mag, $I$~=~14.6~mag;][]{2003AJ....125..984M}.  For
  late-M dwarfs, we have found that the Keck tip-tilt sensor generally
  sees the equivalent of an $(I+1.0)$~mag star.}

On each observing run, we obtained dithered images, offsetting the
telescope by a few arcseconds between every 2--5 images.  The sodium
laser guide star was positioned at the center of the NIRC2
field-of-view for all observations.  The images were reduced in a
standard fashion. We constructed flat fields from the differences of
images of the telescope dome interior with and without continuum lamp
illumination.  Then we created a master sky frame and subtracted it
from the individual images.  For \Lp-band images, the sky subtraction
was done by pairwise subtraction of consecutive images.
Sky-subtracted images were registered and stacked to form a final
mosaic, though all the results described here were based on analysis
of the individual images.  Outlier images with much poorer full-width
half-maxima (FWHM) and/or Strehl ratios were excluded from the
analysis.  We used the publicly available IDL routine
\texttt{NIRC2STREHL}\footnote{\texttt{\url{http://www2.keck.hawaii.edu/optics/lgsao/software/
      nirc2strehl.pro}}} to estimate the Strehl and FWHM at each
epoch, and in Table~\ref{tbl:obs} the mean and standard deviation of
these values over each image set is reported.  The variation in Strehl
ratio and FWHM from between epochs is due to different seeing
conditions.

To determine the relative positions and fluxes of \lhsA\ and \lhsB\ in
the imaging data, we used a simple analytic representation of the PSF
to deblend the two components. The model was the sum of three
elliptical Gaussians in which each Gaussian component was allowed to
have a different FWHM and normalization, but all components had the
same ellipticity and semimajor axis P.A. The best-fit parameters were
found by a Levenberg--Marquardt least-squares minimization in which
all pixels were weighted equally. This fitting procedure yielded a set
of measurements of the projected separation, P.A., and flux ratio for
\lhsbin. We used the astrometric calibration from
\citet{2008ApJ...689.1044G}, with a pixel scale of
9.963$\pm$0.005~mas/pix and an orientation for the detector's
$+y$-axis of $+$0.13$\pm$0.02 east of north. We applied the distortion
correction developed by B. Cameron (private communication, 2007) to
the astrometry, which changed the results well below the 1$\sigma$
level. We also computed the change in the relative astrometry due to
differential chromatic refraction (DCR) in the same manner as
\citep{2009ApJ...692..729D} but did not apply this correction to our
measurements because we found that it induces only a $<$~0.1$\sigma$
change.

\tabletypesize{\scriptsize}
\begin{deluxetable}{lccccc}
\tablecaption{Keck LGS AO Observations \label{tbl:obs}}
\tablewidth{0pt}
\tablehead{
\colhead{Date} &
\colhead{Time} &
\colhead{Airmass} &
\colhead{Filter} &
\colhead{FWHM} &
\colhead{Strehl ratio} \\
\colhead{(UT)} &
\colhead{(UT)} &
\colhead{} &
\colhead{} &
\colhead{(mas)} &
\colhead{}}
\startdata
 2007 Apr 22 & 07:46 & 1.19 &  \Ks\                     &  53.8$\pm$0.2  &  0.277$\pm$0.006  \\
             & 07:56 & 1.19 & $CH_4s$\tablenotemark{\dag} &     \nodata    &      \nodata      \\
             & 10:00 & 1.47 &  \Ks\tablenotemark{\dag}  &     \nodata    &      \nodata      \\
 2008 Jan 15 & 14:48 & 1.21 &  \Ks\tablenotemark{\dag}  &     \nodata    &      \nodata      \\
             & 14:59 & 1.23 &  \Ks\                     &  61.0$\pm$0.8  &  0.201$\pm$0.017  \\
 2009 Jan 23 & 12:56 & 1.21 &  \Ks\                     &  71.2$\pm$1.1  &  0.152$\pm$0.015  \\
             & 13:03 & 1.20 &  $H$                      &  71.9$\pm$2.7  &  0.059$\pm$0.006  \\
             & 13:14 & 1.20 &  $J$                      &  78.4$\pm$4.7  &  0.021$\pm$0.004  \\
             & 13:33 & 1.19 &  \Lp\                     &  93.4$\pm$0.9  &  0.434$\pm$0.069  \\
\enddata
\tablenotetext{\dag}{Aperture masking observations.}
\end{deluxetable}

To assess systematic errors in our PSF-fitting procedure, we also
applied it to simulated Keck images of \lhsbin. We scaled down each
individual image until the primary flux matched the flux from the
faint companion. In this scaled image, the companion flux is at the
level of the noise. We then shifted and added this image to the
original image, matching the binary separation of \lhsbin\ but
avoiding P.A.s within $\pm$60\degs\ of \lhsB, which were masked out in
the original image. After running our PSF-fitting routine on these
images, the resulting scatter in the truth-minus-fitted parameters was
comparable to or somewhat smaller than the rms scatter of the
individual measurements ($\lesssim$~0.5$\sigma$ different), thus
indicating that the rms scatter is a reasonable assessment of the
errors. The Monte Carlo simulations also typically indicated
significant systematic offsets ($\gtrsim$~1$\sigma$) in the astrometry
and also the flux ratio (except for the 2009 data). However, we did
not apply these offsets to any of the parameters derived from PSF
fitting as these offsets have an insignificant impact on the resulting
mass estimate (see Section~\ref{sec:orbit}). They also bring the
\Ks-band flux ratios derived from Keck imaging out of agreement with
the ensemble of flux ratio measurements. This is not unexpected since
the simulated images cannot exactly reproduce the science data, which
has a somewhat asymmetric PSF and speckles that cause distinct
systematic offsets at different separations and P.A.s.

\begin{figure}
\plotone{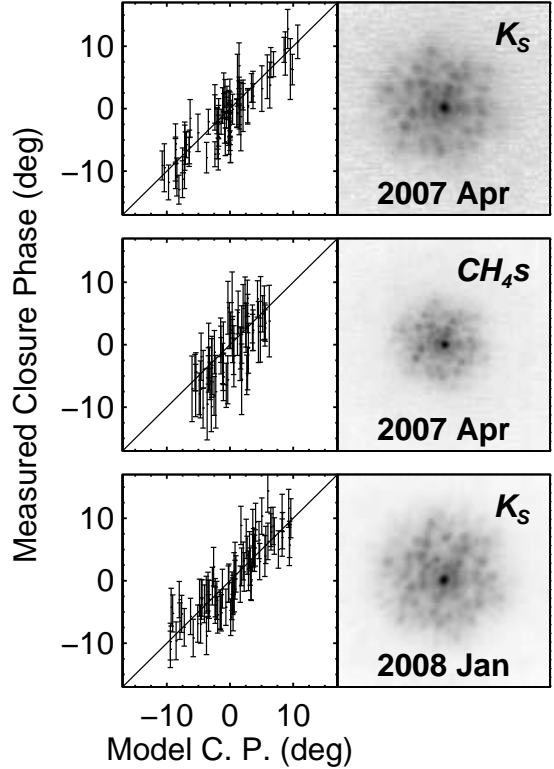}
\caption{ \normalsize Keck interferograms of \lhsbin\ obtained using
  NIRC2's 9-hole aperture mask (square-root stretch).  The left
  panels show the modeled versus measured closure phases (C.P.) at
  each epoch.  The binary parameters were derived from the modeled
  closure phases, and their errors were assessed in a Monte Carlo
  fashion that accounted for the measured closure phase
  errors.  \label{fig:masking}}
\end{figure}

In 2007 and 2008, data were also obtained using the 9-hole
non-redundant aperture mask installed in the filter wheel of NIRC2
\citep{2006SPIE.6272E.103T}. The data were taken in two dither
positions separated by 3$\farcs$5, with four to six 50~s exposures
taken at each dither position. With an integrated $K$-band brightness
of 10.7~mag, \lhsint\ was fainter than typical targets previously
observed with adaptive optics and the aperture mask, but it was still
bright enough not to be limited by readout or background noise.
Typical interferograms are shown in Figure~\ref{fig:masking}. The
pipeline used to reduce the aperture masking data was similar to that
used in previous papers containing NIRC2 masking data
\citep{gl802b-ireland, 2008ApJ...679..762K, 2008ApJ...678L..59I},
except that no comparable single star was observed for calibration of
the closure phases or squared visibilities. For this reason, we chose
to only fit to the closure phases, which from previous experience with
NIRC2 never show an rms scatter of more than 3\degs\ on bright
calibrators in $H$- or $K$-band. Use of squared visibilities would
have required a model of the fringe decorrelation due to an imperfect
AO system, a complexity we chose not to tackle. The resulting closure
phases from our observations are shown in Figure~\ref{fig:masking}.
The closure phases are intrinsically a 4-dimensional function
\citep[the phase of the bispectrum, e.g.,][]{1983ApOpt..22.4028L}, so
they are difficult to present on a 2-dimensional plot. Note that the
closure phases are a linear function of contrast for contrast ratios
larger than $\sim$5:1, so for example if the companion were twice as
bright, then all the measured and model closure phases would be
doubled.

The closure phase uncertainties were initially approximated by the
standard error of the mean calculated from the scatter among
individual exposures. The uncertainties were subsequently increased by
adding a calibration error in quadrature so that the resulting reduced
$\chi^2$ of the fit was 1.0. Although we fit all 84 closure phases
from the 9-hole mask, only 28 of these are formally independent. To
correctly account for this non-diagonal covariance matrix in our
binary fitting, we scaled the errors in the least-squares fit to the
data by $\sqrt{84/28}$. This process has been validated both by a
comparison to fits using full covariance matrices
\citep{2008ApJ...679..762K} and by orbit fits using mixed data that
resulted in a reduced $\chi^2$ consistent with unity, where the orbit
fit had many degrees of freedom \citep[e.g.,][]{2007ApJ...661..496M}.

In Figure~\ref{fig:masking}, it can be seen that binary is clearly
detected at $>$~3$\sigma$ in many individual closure triangles for the
two \Ks-band detections. The $CH_4s$-band detection has only one
closure triangle above 3$\sigma$ but nonetheless has a very clear and
unique fit to all triangles taken together that is consistent with the
other data from the same epoch.

The astrometry and flux ratios derived from the Keck data are given in
Table~\ref{tbl:astrom}. The different data sets at each epoch, both
aperture masking and direct imaging, give consistent measurements of
the binary parameters. When fitting the orbit, we adopt the
measurements from the single data set at each epoch with the smallest
astrometric errors. For the 2008 data, the error in separation is
smaller for the masking data while the P.A. error is smaller in the
imaging data. Because the binary separation is very close to the outer
limit at which masking observations are unambiguous in the 2008 data
($\approx$150~mas; imposed by the smallest non-redundant baselines in
the mask), we conservatively chose to adopt imaging astrometry for the
2008 epoch when fitting the orbit. Regardless of which \Ks-band data
set we use at each epoch, the resulting mass estimate is not
significantly changed, as discussed in Section~\ref{sec:orbit}.

\tabletypesize{\small}
\begin{deluxetable*}{llcccc}
\tablewidth{0pt}
\tablecaption{Best-Fit Binary Parameters for LHS~2397\lowercase{a}AB \label{tbl:astrom}}
\tablehead{
\colhead{Epoch (UT)} &
\colhead{Instrument} &
\colhead{Filter} &
\colhead{$\rho$ (mas)} &
\colhead{P.A. (\degs)} &
\colhead{$\Delta{m}$ (mag)}}
\startdata
 1997 Apr 12  &   \HST/WFPC2-PC1\tablenotemark{*}     &  F814W  &   274$\pm$4   &\phn   87.3$\pm$0.8  &  4.18$\pm$0.08  \\
 2002 Feb  7  &   Gemini/Hokupa`a\tablenotemark{*}    &   \Kp\  &   205$\pm$8   &      152.0$\pm$2.4  &  3.06$\pm$0.45  \\
 2003 May 31  &       VLT/NACO\tablenotemark{*}       &   \Ks\  &   168$\pm$8   &      188.6$\pm$1.2  &  2.75$\pm$0.16  \\
 2006 Jan 12  &       VLT/NACO\tablenotemark{*}       &   \Ks\  &   129$\pm$5   &      276.2$\pm$1.4  &  2.84$\pm$0.16  \\
 2007 Apr 22  & Keck/NIRC2                            &   \Ks\  & 119.7$\pm$2.9 &      348.4$\pm$0.8  &  2.94$\pm$0.10  \\
              & Keck/NIRC2 (masking)                  & $CH_4s$ & 112.5$\pm$2.8 &      347.0$\pm$1.3  &  3.37$\pm$0.24  \\
              & Keck/NIRC2 (masking)\tablenotemark{*} &   \Ks\  & 115.3$\pm$1.8 &      349.3$\pm$1.1  &  2.78$\pm$0.11  \\
 2008 Jan 15  & Keck/NIRC2 (masking)                  &   \Ks\  & 144.6$\pm$2.3 &\phn   24.1$\pm$0.7  &  2.97$\pm$0.11  \\
              & Keck/NIRC2\tablenotemark{*}           &   \Ks\  &   146$\pm$4   &\phn   25.1$\pm$0.5  &  2.70$\pm$0.12  \\
 2009 Jan 23  & Keck/NIRC2\tablenotemark{*}           &   \Ks\  & 204.0$\pm$1.5 &\phn   53.8$\pm$0.3  &  2.80$\pm$0.03  \\
              & Keck/NIRC2                            &   $H$   & 202.0$\pm$2.4 &\phn   53.9$\pm$0.5  &  2.96$\pm$0.05  \\
              & Keck/NIRC2                            &   $J$   &   196$\pm$5   &\phn   53.9$\pm$0.8  &  3.12$\pm$0.08  \\
              & Keck/NIRC2                            &   \Lp\  &   200$\pm$3   &\phn   53.3$\pm$0.4  &  1.92$\pm$0.06  \\
\enddata
\tablenotetext{*}{Used in the orbit fit.}
\end{deluxetable*}

\subsection{Gemini/Hokupa`a \label{sec:gem}}

LHS~2397aAB was imaged on UT 2002 February 7 by the Hokupa`a curvature
AO system at the Gemini-North Telescope on Mauna Kea, Hawai`i. We
retrieved these raw data from the Gemini science archive and
registered, sky-subtracted, and performed cosmic ray rejection on the
images. Figure~\ref{fig:data} shows a typical image from of one of the
15 \Kp-band 3~s exposures which were used to derive the astrometry for
\lhsbin. Analysis of these data has previously been presented by
\citet{2003ApJ...584..453F}; however, astrometric errors were not
derived in that work, so we conducted our own analysis of these data.

We used the same analytic PSF-fitting routine as for the Keck data to
fit the Gemini images of \lhsbin. Adopting the nominal instrument
pixelscale of 19.98$\pm$0.08~mas/pix, we found a separation of
205$\pm$8~mas, where the uncertainty is the standard deviation of
measurements from individual dithers. This is in good agreement with
the 207~mas separation reported by \citet{2003ApJ...584..453F}.
However, the P.A. we find (206.8$\pm$2.4\degs) is significantly
different from the 152.0\degs\ P.A. reported by
\citet{2003ApJ...584..453F}. This must be due to an inaccurate
reporting of the orientation of the camera in the header of the
archival data, which indicates that the P.A. of the $+y$-axis is
0\degs. Such a discrepancy was previously seen in our work with
archival Gemini/Hokupa`a data of \hdbin\ and Gl~569Bab
\citep{2009ApJ...692..729D}. Thus, we adopt the published value of the
P.A., but not its quoted error, for our analysis.

To assess systematic errors in the Gemini astrometry, we simulated
many Gemini images of \lhsbin\ in the same fashion as the Keck data
and fit each one with our analytic PSF model. The resulting scatter in
the truth-minus-fitted parameters was 14~mas in separation and 4\degs\
in P.A., with no significant systematic offsets. These errors are much
larger than the rms scatter of the individual measurements, which is
likely because the PSF in the science data is somewhat asymmetric so
that the offset between the best-fit and input parameters varies
widely, depending on the binary P.A., and by averaging over many
binary P.A.s we overestimate the scatter. However, as shown in
Section~\ref{sec:orbit}, whether we adopt the rms scatter or the Monte
Carlo errors only affects the $\chi^2$ of the orbit and not the
resulting mass estimate. In fact, even if the Gemini epoch is excluded
from the orbit fit the resulting mass estimate changes
insignificantly.

\subsection{VLT/NACO \label{sec:vlt}}

We retrieved archival images of \lhsbin\ obtained with the Very Large
Telescope (VLT) at Paranal Observatory on UT 2003 May 31 and 2006 Jan
15.  These data were taken with the NACO adaptive optics system
\citep{2003SPIE.4841..944L, 2003SPIE.4839..140R} using the N90C10
dichroic at both epochs and the S27 and S13 cameras in 2003 and 2006,
respectively.  The nominal pixel scales of these cameras are
27.053$\pm$0.019~mas/pix and
13.221$\pm$0.017~mas/pix.\footnote{\texttt{\url{http://www.eso.org/sci/facilities/paranal/instruments/
      naco/doc/VLT-MAN-ESO-14200-2761\_v83.3.pdf}}} We registered,
sky-subtracted, and performed cosmic ray rejection on the raw archival
images.  The 2003 data comprise five \Ks-band 5~s exposures, and the
2006 data comprise 32 \Ks-band 10~s exposures.  Typical images from
each data set are shown in Figure~\ref{fig:data}.

We used the same analytic PSF-fitting routine as for the Keck data to
fit the VLT images of \lhsbin. From the 2003 VLT data we derived a
separation of 168$\pm$8~mas, a P.A. of 188.6$\pm$1.2\degs, and a
\Ks-band flux ratio of 2.75$\pm$0.16~mag. From the 2006 VLT data we
derived a separation of 129$\pm$5~mas, a P.A. of 276.2$\pm$1.4\degs,
and a \Ks-band flux ratio of 2.84$\pm$0.16~mag. The quoted uncertainty
is the standard deviation of measurements from individual dithers. To
assess any additional systematic errors in the VLT astrometry, we
simulated many VLT images of \lhsbin\ in the same fashion as the
Gemini and Keck data. These simulations yielded equivalent errors to
the rms scatter quoted above ($\lesssim$~0.2$\sigma$ different), and
thus the quoted errors are a good representation of the uncertainties.

\begin{figure}[b]
\plotone{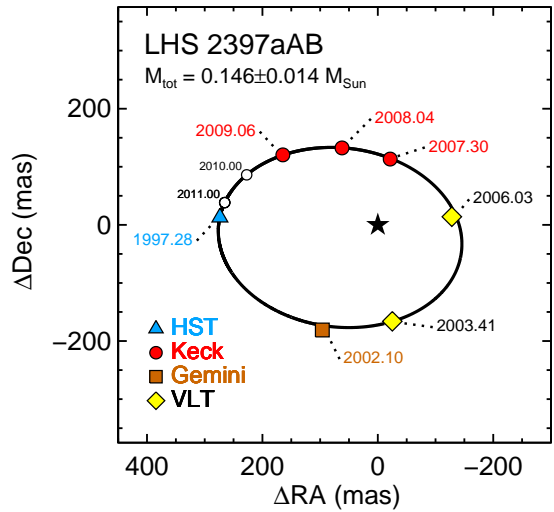}
\caption{\normalsize Relative astrometry for \lhsbin\ along with the
  best fitting orbit (reduced $\chi^2$ of 1.04 for 7 degrees of
  freedom). The empty circles show the predicted location of \lhsB\ in
  the future.  Error bars are comparable to or smaller than the
  plotting symbols. The orbit is very well constrained, so the
  uncertainty in the total mass is dominated by the 3.0\% parallax
  error. \label{fig:orbit-skyplot}}
\end{figure}

\begin{figure*}[ht]
\plotone{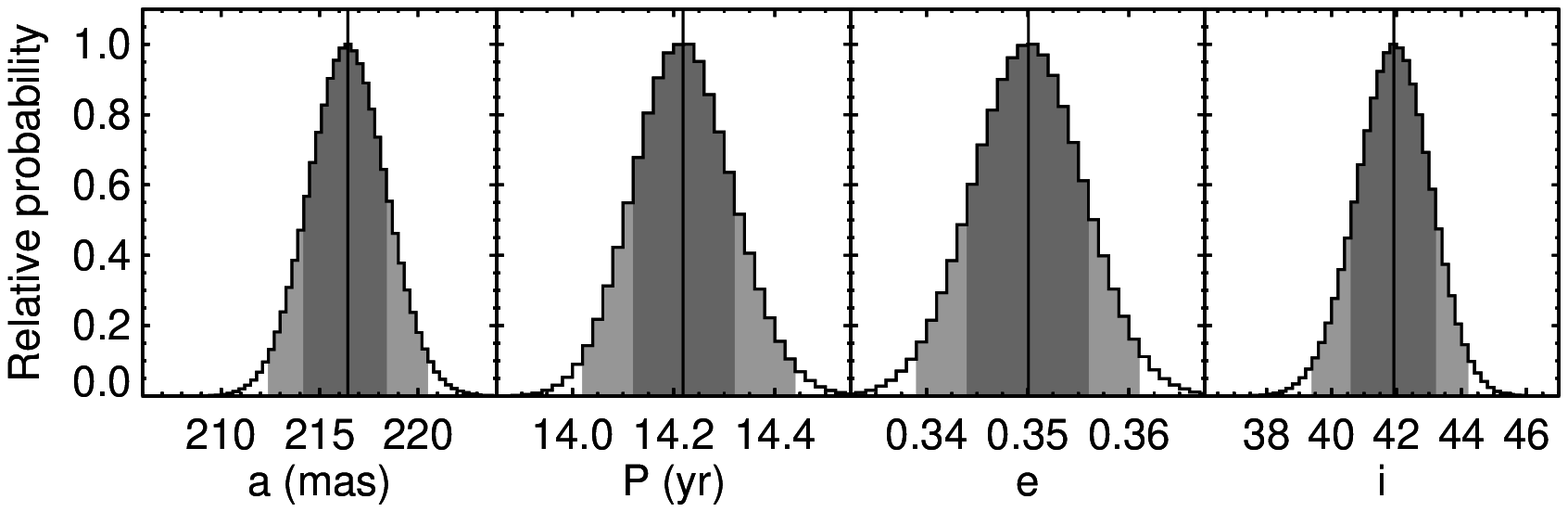} \\
\plotone{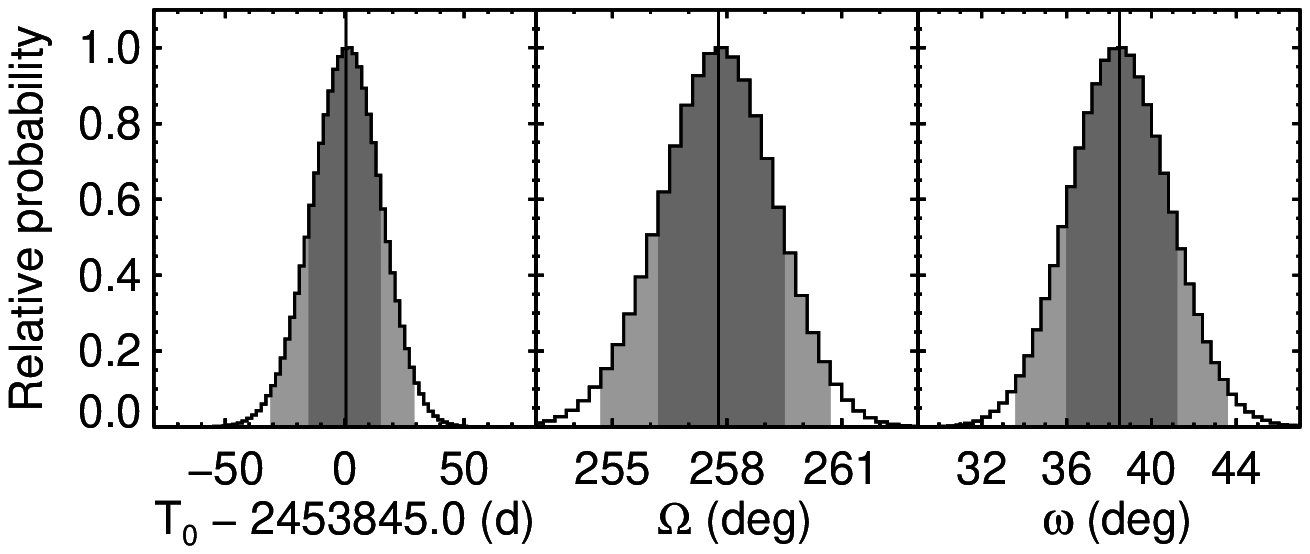}
\caption{\normalsize Probability distributions of all orbital
  parameters derived from the MCMC analysis: semimajor axis ($a$),
  orbital period ($P$), eccentricity ($e$), inclination ($i$), epoch
  of periastron ($T_0$), P.A. of the ascending node ($\Omega$), and
  argument of periastron ($\omega$).  Each histogram is shaded to
  indicate the 68.3\% and 95.4\% confidence regions, which correspond
  to 1$\sigma$ and 2$\sigma$ for a normal distribution, and the solid
  vertical lines represent the median values.  Note that $T_0$ is
  shown in days since UT 2006 Apr 19 12:00 for
  clarity. \label{fig:orbit-parms}}
\end{figure*}

\begin{figure}[b]
\plotone{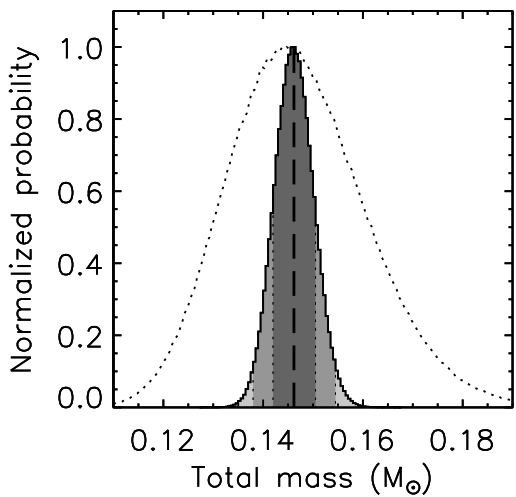}
\caption{\normalsize Probability distribution of the total mass of
  \lhsbin\ resulting from our MCMC analysis.  The histogram is shaded
  to indicate the 68.3\%, 95.4\%, and 99.7\% confidence regions, which
  correspond to 1$\sigma$, 2$\sigma$, and 3$\sigma$ for a normal
  distribution. The dashed line represents the median value of
  0.146~\Msun. The standard deviation of the distribution is
  0.004~\Msun. The dotted unshaded curve shows the final mass
  distribution after accounting for the additional 9.0\% error due to
  the uncertainty in the parallax; the result is essentially Gaussian.
  The confidence limits for both distributions are given in
  Table~\ref{tbl:orbit}.  \label{fig:orbit-mass}}
\end{figure}


\section{Results \label{sec:results}}

\subsection{Orbit Determination \& Dynamical Mass \label{sec:orbit}}

The orbit of \lhsbin\ is very well constrained as our observations
cover most of the orbital period. In order to search for the influence
of parameter degeneracies in our orbit fit and determine robust
confidence limits on the orbital parameters, we used a Markov Chain
Monte Carlo (MCMC) technique \citep[e.g.,][]{bremaud99:markov_chain}
for orbit fitting, in addition to a gradient descent technique. In
short, the MCMC method constructs a series of steps through the model
parameter space such that the resulting set of values (the ``chain'')
is asymptotically equivalent to the posterior probability distribution
of the parameters being sought. The code that performed the MCMC fit
is described in detail in the study of \twomassbin\ by
\citet{2008ApJ...689..436L}. Chains all had lengths of 2$\times$10$^8$
steps, and the correlation length of our most correlated chain, as
defined by \citet{2004PhRvD..69j3501T}, was 230 for the argument of
periastron. This gives an effective length of the chain of
8.7$\times$10$^5$, which in turn gives statistical uncertainties in
the parameter errors of about $1/\sqrt{8.7\times10^5}$~=~0.11\%, i.e.,
negligible.

We used uniform priors in period ($P$), semimajor axis ($a$), P.A. of
the ascending node ($\Omega$), argument of periastron ($\omega$), and
time of periastron passage ($T_0$).  We used a prior in inclination
proportional to $\sin(i)$ (i.e., random orbital orientation) and an
eccentricity prior of $f(e)$~=~2$e$ \citep[e.g.,
see][]{1991A&A...248..485D}.  Figure~\ref{fig:orbit-parms} shows the
resulting MCMC probability distributions for the seven orbital
parameters of \lhsbin.  The best-fit parameters and their confidence
limits are given in Table~\ref{tbl:orbit}, and the best-fit orbit is
shown in Figure~\ref{fig:orbit-skyplot}.  The reduced $\chi^2$ of the
orbital solution is 1.04, with 7 degrees of freedom.

Applying Kepler's Third Law to the period and semimajor axis
distributions gives the posterior probability distribution for the
total mass of \lhsbin, with a median of 0.146~\Msun, a standard
deviation of 0.004~\Msun, and 68.3(95.4)\% confidence limits of
$^{+0.004}_{-0.004}$($^{+0.008}_{-0.008}$)~\Msun\
(Figure~\ref{fig:orbit-mass}). The MCMC probability distribution of
the total mass does not include the uncertainty in the parallax
(3.0\%), which by simple propagation of errors would contribute an
additional 9.0\% uncertainty in mass. In fact, since the MCMC-derived
mass distribution is slightly asymmetric, we account for this
additional error by randomly drawing a normally distributed parallax
value for each step in the chain, which we then used to compute the
total mass. The resulting mass distribution is nearly
indistinguishable from Gaussian (Figure~\ref{fig:orbit-mass}). Our
final determination of the total mass is
0.146$^{+0.015}_{-0.013}$($^{+0.031}_{-0.024}$)~\Msun\ at 68.3(95.4)\%
confidence.

\tabletypesize{\small}
\begin{deluxetable*}{lcccc}[ht]
\tablewidth{0pt}
\tablecaption{Derived Orbital Parameters for LHS~2397\lowercase{a}AB \label{tbl:orbit}}
\tablehead{
\colhead{}   &
\multicolumn{3}{c}{MCMC}      &
\colhead{\orbit\tablenotemark{\dag}} \\
\cline{2-4}
\colhead{Parameter}   &
\colhead{Median}      &
\colhead{68.3\% c.l.} &
\colhead{95.4\% c.l.} &
\colhead{} }
\startdata
Semimajor axis $a$ (mas)                                         &  216.4 &    $-$1.9, 2.0    &   $-$3.8, 3.9     &  216.5$\pm$2.0\phn\phn    \\
Orbital period $P$ (yr)                                          &  14.22 &   $-$0.10, 0.10   &  $-$0.19, 0.20    &  14.22$\pm$0.10\phn       \\
Eccentricity $e$                                                 &  0.350 &  $-$0.005, 0.005  & $-$0.011, 0.011   &  0.345$\pm$0.005          \\
Inclination $i$ (\degs)                                          &  41.9  &    $-$1.1, 1.1    &   $-$2.3, 2.2     &   42.0$\pm$1.1\phn        \\
Time of periastron passage $T_0-2453845.0$\tablenotemark{a} (JD) &    0   &     $-$15, 14     &    $-$30, 28      &  \phn0$\pm$15             \\
P.A. of the ascending node $\Omega$ (\degs)                      &  257.8 &    $-$1.5, 1.5    &   $-$3.1, 2.9     &  257.8$\pm$1.5\phn\phn    \\
Argument of periastron $\omega$ (\degs)                          &   38.5 &    $-$2.3, 2.4    &   $-$4.6, 4.8     &   38.5$\pm$2.4\phn        \\
Total mass (\Msun): fitted\tablenotemark{b}                      & 0.146  &  $-$0.004, 0.004  & $-$0.008, 0.008   &  0.146$\pm$0.004          \\
Total mass (\Msun): final\tablenotemark{c}                       & 0.146  &  $-$0.013, 0.015  & $-$0.024, 0.031   &  0.146$\pm$0.14\phn       \\
Reduced $\chi^2$ (7 degrees of freedom)                          &  1.04  &       \nodata     &       \nodata     &        1.04               \\
\enddata

\tablenotetext{\dag}{The orbital parameters determined by \orbit with
  their linearized 1$\sigma$ errors.}

\tablenotetext{a}{UT 2006 Apr 19 12:00:00.0} 

\tablenotetext{b}{The ``fitted'' total mass represents the direct
  results from fitting the observed orbital motion of the two
  components without accounting for the parallax error.  For the
  linearized \orbit\ error, the covariance between $P$ and $a$ is
  taken into account.}

\tablenotetext{c}{The ``final'' total mass includes the additional
  9.0\% error in the mass due to the error in the parallax.  This
  final mass distribution is essentially Gaussian.}

\end{deluxetable*}

As an independent verification of our MCMC results, we also fit the
orbit of \lhsbin\ using the linearized least-squares routine \orbit\
\citep[described in][]{1999A&A...351..619F}.  All of the orbital
parameters and their errors are consistent between the \orbit\ and
MCMC results, as expected for well constrained parameters with nearly
Gaussian probability distributions.  The $\chi^2$ and total mass of
the \orbit\ solution and the MCMC solution were identical.

From the visual orbit alone there is a 180\degs\ ambiguity in the
P.A. of the ascending node ($\Omega$), and correspondingly $\omega$,
that can only be resolved by radial velocity measurements.
\citet{2006AJ....132..663B} measured the radial velocity of \lhsint\
at two epochs (UT 1995 Mar 12 and 2002 May 19) in the optical where
\lhsB\ is essentially invisible.  For a mass ratio of 0.7
(Section~\ref{sec:qratio}), the expected difference in the radial
velocity of the primary between these two epochs is $\pm$1.8~\kms,
where the unknown sign reflects the 180\degs\ ambiguity in $\Omega$.
\citet{2006AJ....132..663B} measured a velocity difference of
$+$1.1$\pm$1.6~\kms, which is consistent with $+$1.8~\kms\ but
discrepant with $-$1.8~\kms\ at 1.8$\sigma$.

We have tested whether varying the input astrometry and corresponding
uncertainties affects the orbit-fitting results using \orbit.
Regardless of what astrometry was used, the mass estimate and its
uncertainty did not change significantly. We tested many permutations
using \orbit, such as using Keck imaging instead of masking astrometry
or Monte Carlo instead of rms errors, and we found that the resulting
mass estimates ranged from 0.142--0.150~\Msun\ and uncertainties
ranged from 0.003--0.006~\Msun. The $\chi^2$ of these different
permutations ranged from 5--17 (7 degrees of freedom), so many of
these scenarios are clearly not as favorable as our default solution
that has a $\chi^2$ of 7.27. The permutations with lower $\chi^2$
simply correspond to the cases with somewhat larger assumed
astrometric uncertainties, and these had essentially identical masses
and mass errors as our default solution.

In fact, any single epoch can be excluded and the resulting dynamical
mass changes by $<$~0.006~\Msun, with the uncertainty from the orbit
alone increasing to only $\pm$0.006~\Msun, which is well below the
error due to the parallax uncertainty. Furthermore, pairs of epochs
can even be excluded without significantly changing the dynamical
mass. For example, both of the \HST\ and Gemini epochs or both of the
VLT epochs can be excluded and the resulting dynamical mass changes by
$<$~0.008~\Msun\ and the error from the orbit alone increases to
0.013~\Msun\ (i.e., comparable to the error due to the parallax
uncertainty). Thus, our mass estimate for \lhsbin\ and its uncertainty
is insensitive to the input astrometry and the method of determining
the measurement errors.

\begin{figure*}
\plotone{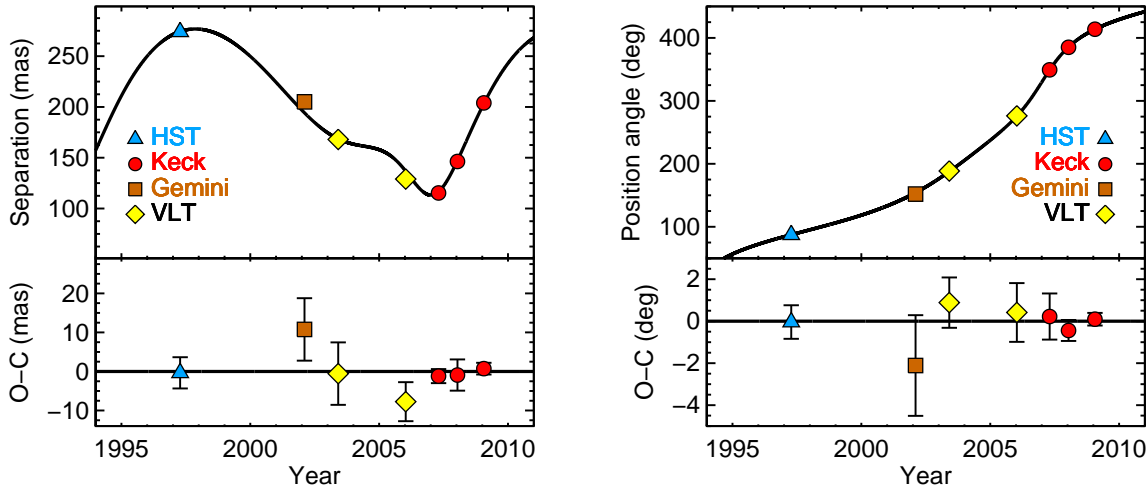}
\caption{\normalsize Measurements of the projected separation ({\em
    left}) and P.A. ({\em right}) of \lhsbin.  The best-fit orbit is
  shown as a solid line.  The bottom panels show the observed minus
  predicted measurements with observational error bars.  The fact that
  the 2007 Keck astrometry shows the smallest errors even though the
  data were obtained near the binary's minimum separation highlights
  the exquisite precision that can be obtained for such a tight,
  high-contrast binary using aperture
  masking. \label{fig:orbit-sep-pa}}
\end{figure*}

\subsection{Spectral Types \label{sec:spt}}

Spectral type determination for the components of \lhsbin\ is impeded
by the lack of resolved spectroscopy. However, the optical
classification of M8 \citep{1995AJ....109..797K} can be readily
applied to the primary since the $I$-band flux ratio of
$\approx$~4~mag indicates that \lhsB\ contributes negligibly to the
flux in the unresolved optical spectrum. The spectral type of \lhsB\
must be estimated indirectly, and we have done so in a variety of
ways.

The available $IJHK\Lp$ colors of \lhsB\ do not provide a strong
constraint on the spectral type; however, they are consistent with
late-L dwarfs \citep{2002AJ....124.1170D, gol04, 2006PASP..118..659L}.
Unfortunately, even if the colors of \lhsB\ were measured perfectly,
the dispersion among known L dwarfs limits a spectral type estimate
from colors alone to at least $\pm$2 subclasses.

We also used the empirical absolute magnitude relations of
\citet{2006astro.ph..5037L}\footnote{These empirical relations are
  based on near-infrared spectral types.} with known and suspected
binaries removed to estimate the spectral type of \lhsB. We converted
these relations from the MKO to the 2MASS photometric system using the
relations of \citet{2004PASP..116....9S}. In addition, we derived the
\Lp-band absolute magnitude relation for the same sample, though 7 of
the 29 objects from \citet{2006astro.ph..5037L} could not be used as
they did not have published \Lp-band photometry
\citep{2002ApJ...564..452L, gol04}. The resulting polynomial fit was
\begin{eqnarray}
  M_{L^{\prime}} & = & 9.759~+~0.1507\times{\rm SpT}~+~2.868\times10^{-2}\times{\rm SpT}^2 \nonumber \\
               &   & -~3.674\times10^{-3}\times{\rm SpT}^3~+~1.338\times10^{-4}\times{\rm SpT}^4 \nonumber
\end{eqnarray}
with an rms scatter about this relation of 0.18~mag. By simply
interpolating these absolute magnitude relations, the $K$-band
photometry of \lhsB\ gave a spectral type of L7.5$\pm$0.3, the
$H$-band photometry gave L6.6$\pm$0.3, and the \Lp-band photometry
gave L7.4$\pm$0.5. These quoted uncertainties only account for the
photometric errors (i.e., the 1$\sigma$ spectral type range
corresponds to the 1$\sigma$ range in photometry) and not scatter in
the empirical relations. Because of the flattening of the absolute
magnitude relation in $J$-band, we were unable to estimate a robust
spectral type from the $J$-band photometry, although it is consistent
with the spectral type ranges listed above.

\begin{figure}[hb]
\plotone{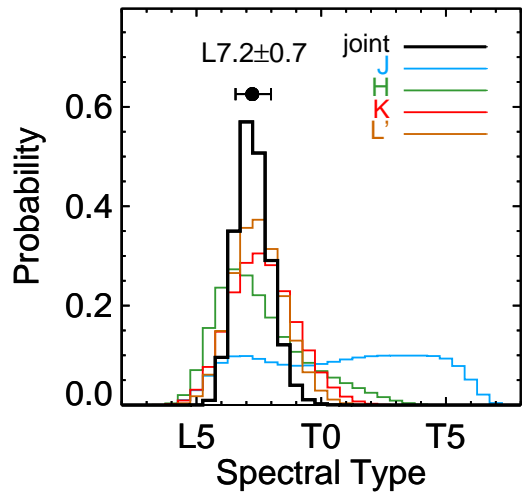}
\caption{\normalsize The final probability distribution of the
  spectral type of \lhsB\ (solid black line), derived using the
  absolute magnitude relations of \citet{2006astro.ph..5037L}, with
  suspected binaries removed, and our own \Lp-band relation as
  described in the text. The joint probability distribution of the
  spectral type was derived from $J$-band, $H$-band, $K$-band, and
  \Lp-band photometry (thin colored lines). The filled circle with
  error bars denotes the median and 1$\sigma$ confidence limits of
  this final spectral type distribution. \label{fig:spt}}
\end{figure}

We also employed a more sophisticated method for spectral type
estimation that accounts for both the uncertainty in the absolute
magnitude of \lhsB\ and the scatter in the empirical relations. At
each half-integer spectral type, we considered two Gaussian
probability distributions: (1) the distribution of the observed
absolute magnitude of \lhsB\ and (2) the distribution of absolute
magnitudes predicted by the empirical relation from the overall
scatter in the relation as given by \citet{2006astro.ph..5037L}. We
multiplied these two distributions and integrated the result in order
to obtain a measure of the relative probability of \lhsB\ having that
spectral type. If the two distributions overlapped substantially, then
the integral was large, but if the two distributions were very
discrepant the integral went to zero. Thus, we built an ensemble of
relative probabilities which we adopted as the probability
distribution of the spectral type of \lhsB\ (Figure~\ref{fig:spt}). We
did this separately for each bandpass, and the derived spectral type
and uncertainty was T0.9$^{+3.5}_{-4.0}$ for $J$-band,
L7.2$^{+2.2}_{-1.3}$ for $H$-band, L7.6$^{+1.4}_{-1.3}$ for $K$-band,
and L7.4$^{+1.1}_{-1.0}$ for \Lp-band (68.3\% confidence limits).
Since each bandpass represents an independent constraint on the
spectral type, we took the product of these different spectral type
distributions to determine the final spectral type probability
distribution (Figure~\ref{fig:spt}). This is conceptually equivalent
to taking the weighted mean of the spectral types listed above, and
the result is L7.2$\pm$0.7.\footnote{If we use
  the \citet{2006astro.ph..5037L} empirical relations which \emph{do
    not} exclude suspected binaries, the resulting spectral type of
  \lhsB\ is L7.6$^{+1.1}_{-0.9}$.} Because of the good agreement
between the spectral types inferred from different bandpasses, the
final derived spectral type of \lhsB\ is essentially the same with
smaller errors.

We adopt the best spectral type estimates available: for \lhsA\ this
is M8$\pm$0.5 from the optical spectrum, and for \lhsB\ we assign a
spectral type of L7$\pm$1 from the absolute magnitude relations, where
the uncertainty accounts for both the observational errors in the
photometry and the scatter in the empirical relations.

\begin{deluxetable}{lccc}
\tablewidth{0pt}
\tablecaption{Measured Properties of LHS~2397\lowercase{a}AB \label{tbl:meas}}
\tablehead{
\colhead{Property\tablenotemark{a}}   &
\colhead{\lhsA} &
\colhead{\lhsB} &
\colhead{Ref.}       }
\startdata
\Mtot\ (\Msun)             & \multicolumn{2}{c}{   0.146$\pm$0.014  }  &   1   \\
Semimajor axis (AU)        & \multicolumn{2}{c}{    3.09$\pm$0.10   }  &  1,2  \\
$d$ (pc)                   & \multicolumn{2}{c}{    14.3$\pm$ 0.4\phn} &   2   \\
Spectral Type              &\phs\phd  M8$\pm$0.5   &\phs      L7$\pm$1         &  1,3  \\
$J$ (mag)                  &\phs   12.11$\pm$0.02 &\phs   15.23$\pm$0.08      &  1,4  \\
$H$ (mag)                  &\phs   11.26$\pm$0.02 &\phs   14.22$\pm$0.05      &  1,4  \\
$K$ (mag)                  &\phs   10.83$\pm$0.02 &\phs   13.64$\pm$0.04      &  1,4  \\
\Lp\ (mag)                 &\phs   10.20$\pm$0.02 &\phs   12.12$\pm$0.05      &  1,5  \\
$J-H$ (mag)                &\phs\phn0.85$\pm$0.03 &\phs\phn1.01$\pm$0.10      &  1,4  \\
$H-K$ (mag)                &\phs\phn0.43$\pm$0.03 &\phs\phn0.58$\pm$0.07      &  1,4  \\
$J-K$ (mag)                &\phs\phn1.28$\pm$0.03 &\phs\phn1.59$\pm$0.09      &  1,4  \\
$K-\Lp$ (mag)              &\phs\phn0.63$\pm$0.03 &\phs\phn1.51$\pm$0.07      & 1,4,5 \\
$M_V$ (mag)                &\phs   18.80$\pm$0.07  &\phs       \nodata         &   2   \\
$M_J$ (mag)                & \phs  11.34$\pm$0.07  &\phs   14.46$\pm$0.10      & 1,2,4 \\
$M_H$ (mag)                & \phs  10.49$\pm$0.07  &\phs   13.45$\pm$0.08      & 1,2,4 \\
$M_K$ (mag)                & \phs  10.06$\pm$0.07  &\phs   12.87$\pm$0.08      & 1,2,4 \\
$M_{L^{\prime}}$ (mag)       &\phs\phn9.43$\pm$0.07  &\phs   11.35$\pm$0.09      & 1,2,5 \\
$\log$(\Lbol/\Lsun)        &\phn $-$3.42$\pm$0.03  &\phn $-$4.52$\pm$0.03      &   1   \\
$\Delta\log$(\Lbol)        &   \multicolumn{2}{c}{ 1.102$\pm$0.011 }   &   1   \\
\enddata

\tablenotetext{a}{All near-infrared photometry on the MKO system.}

\tablerefs{(1)~This work; (2)~\citet{1992AJ....103..638M};
  (3)~\citet{1995AJ....109..797K}; (4)~\citet{2mass};
  (5)~\citet{2002ApJ...564..452L}.}

\end{deluxetable}

\begin{figure}[t]
\plotone{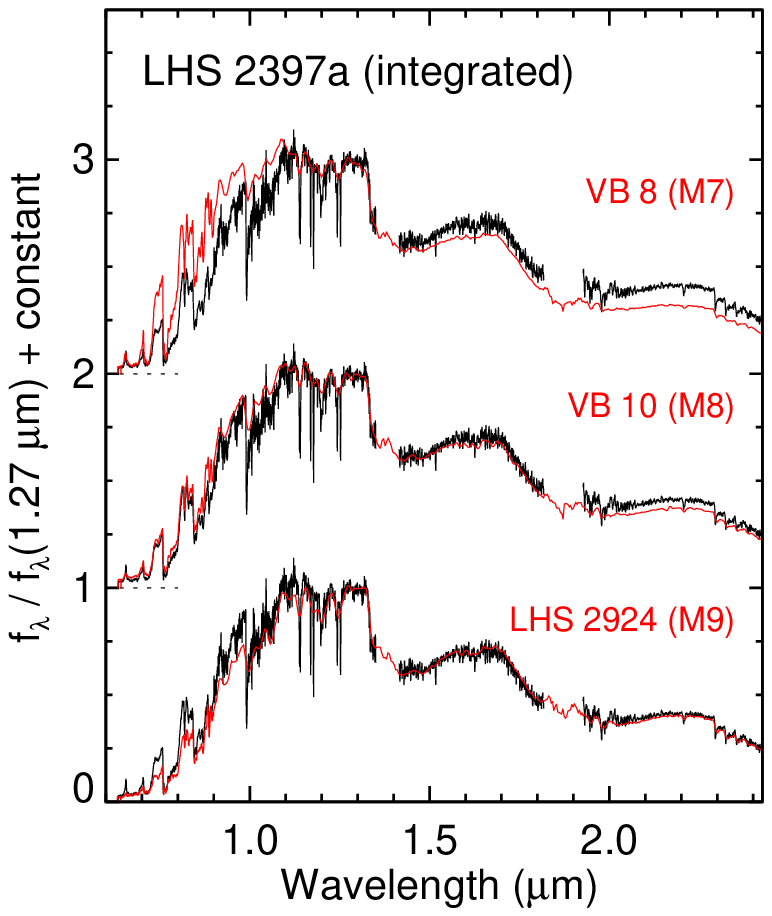}
\caption{\normalsize The integrated light spectrum of \lhsint, with
  optical data ($<$~0.9~\micron) taken from
  \citet{1995AJ....109..797K} and our near-infrared data obtained with
  IRTF/SpeX. Data for M~dwarf spectral standards are shown for
  comparison \citep{1990ApJ...354L..29H, 2004AJ....127.2856B,
    2006AJ....131.1007B, 2008ApJ...681..579B}.  \lhsint\ is optically
  typed as M8, and its near-infrared ($H$ and $K$ band) spectrum is
  discrepant with the standard VB~10 due to additional flux from the
  very red L7 dwarf companion \lhsB. \label{fig:spectra}}
\end{figure}

\subsection{Bolometric Luminosities \label{sec:lbol}}

We computed bolometric luminosities for \lhsbin\ in two different
ways: (1)~directly using integrated-light measurements of \lhsint\ and
(2)~using our resolved photometry for the individual binary components
along with spectral templates. \citet{1995AJ....109..797K} measured
the optical integrated -light spectrum of \lhsint\
($\approx$0.64--0.92~\micron).\footnote{Publicly available at
  \texttt{\url{http://www.dwarfarchives.org}}.}  On 2008 June 27 UT we
obtained a near-infrared spectrum of \lhsint\
($\approx$0.82--2.4~\micron) using IRTF/SpeX
\citep{1998SPIE.3354..468R} in SXD mode ($R$~=~1200), which we reduced
using the SpeXtool software package \citep{2003PASP..115..389V,
  2004PASP..116..362C}.  We combined these spectra, using optical data
below 0.9~\micron\ and our SpeX data above 0.9~\micron.
Figure~\ref{fig:spectra} shows the resulting integrated-light spectrum
of \lhsint, along with M~dwarf spectral standards from
\citet{1991ApJS...77..417K}.\footnote{We combined optical and
  near-infrared data for the M~dwarf spectral standards in the same
  way as \lhsint.  The optical data are from
  \citet{1990ApJ...354L..29H}, and the near-infrared data are SpeX
  prism spectra from \citet{2004AJ....127.2856B},
  \citet{2006AJ....131.1007B}, and \citet{2008ApJ...681..579B}.}

\begin{figure*}[ht]
\plotone{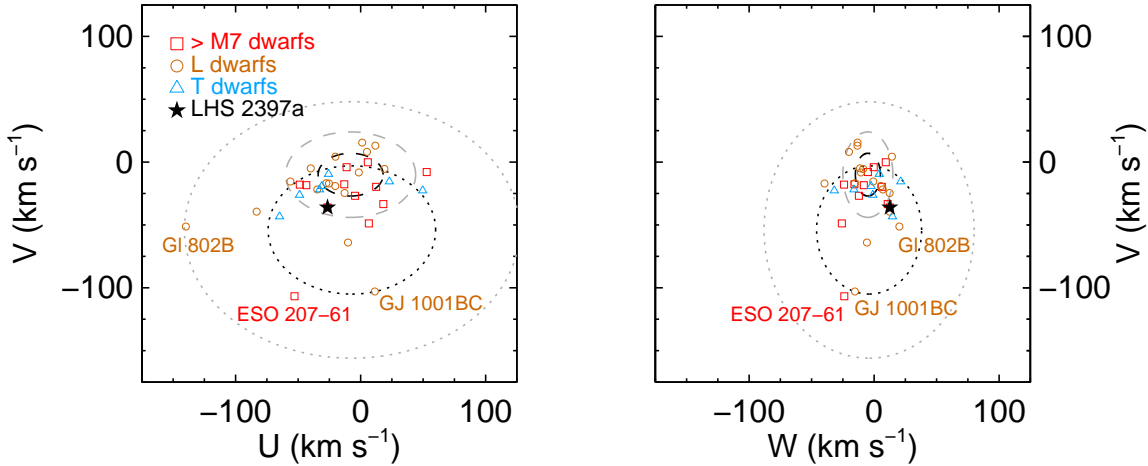}
\caption{ \normalsize The heliocentric space velocity of \lhsint\
  (star) is shown alongside other ultracool dwarfs: $>$~M7 dwarfs
  (squares), L dwarfs (circles), and T dwarfs (triangles). The
  1$\sigma$ and 2$\sigma$ ellipsoids of the thin disk (dashed) and
  thick disk (dotted) as predicted by the Besan\c{c}on galaxy model
  \citep{2003A&A...409..523R} are also shown for comparison. The space
  velocity of \lhsint\ is consistent with other ultracool dwarfs, and
  we derive a 99.2\% thin disk membership
  probability. \label{fig:uvw}}
\end{figure*}

We derived a total bolometric luminosity of
$\log(\Lbol/\Lsun)$~=~$-$3.33$\pm$0.03 for \lhsint\ using the
integrated-light spectrum (0.6--2.4~\micron), \Lp-band (3.8~\micron)
photometry from \citet{2002ApJ...564..452L}, and 24~\micron\
\Spitzer/MIPS photometry from \citet{2007ApJ...667..527G}.  We
neglected any flux at shorter wavelengths, interpolated between gaps
in the data, and extrapolated the flux beyond 24~\micron\ assuming a
blackbody.  We determined the luminosity error in a Monte Carlo
fashion by adding randomly drawn noise to our data over many trials
and computing the rms of the resulting luminosities.

For \lhsint, the SpeX spectrum and \Lp-band photometry alone account
for 98\% of its luminosity.  Thus, it is possible to accurately
estimate the luminosities of the individal components using only our
resolved $JHK\Lp$ photometry and spectral templates to interpolate
properly between bandpasses.  We selected objects from the SpeX Prism
Libraries\footnote{\texttt{http://www.browndwarfs.org/spexprism/}}
that had similar colors to \lhsA\ and \lhsB\ and calibrated them to
match our resolved $JHK$ photometry.  In fact, we found that the exact
choice of the templates had an insignificant impact on the resulting
luminosity measurements.  We integrated over the template spectra,
neglecting flux at shorter wavelengths, interpolating the flux between
the spectra and \Lp\ band, and extrapolating to longer wavelengths
assuming a blackbody.  Thus, we derived luminosities of
$\log(\Lbol/\Lsun)$~=~$-$3.42$\pm$0.03 for \lhsA\ and $-$4.52$\pm$0.03
for \lhsB.  The sum of these luminosities is $-$3.39$\pm$0.03, which
is consistent with the \Lbol\ derived from integrated-light data.  The
dominant source of error in these luminosities is the parallax, which
is common to both components, and in the following analysis we
correctly account for this covariance.

\subsection{Age Constraints from Kinematics and Activity
  \label{sec:age}}

The space motion of \lhsint\ could provide a constraint on its age,
especially if it were found to be kinematically associated with a
population of known age (e.g., the thick disk).  Using the proper
motion and parallax from \citet{1992AJ....103..638M} and a radial
velocity of 34$\pm$2~\kms\ \citep{2003ApJ...583..451M,
  1998MNRAS.301.1031T}, we computed the heliocentric velocity of
\lhsint\ to be $(U, V, W)$~=~($-$26.5$\pm$0.8, $-$36.1$\pm$1.5,
$+$12.5$\pm$1.4)~\kms.\footnote{This is consistent with the
  heliocentric velocity derived by \citet{2003ApJ...584..453F}, but
  inconsistent with the velocity computed by
  \citet{1998MNRAS.301.1031T}.  It appears that their space velocity
  for \lhsint\ is discrepant by $\approx$2$\times(U, V, W)_{\sun}$ as
  a result of subtracting rather than adding the solar motion to the
  heliocentric velocities.}  We adopt the sign convention for $U$ that
is positive toward the Galactic center and account for the
errors in the parallax, proper motion, and radial velocity in a Monte
Carlo fashion.

For comparison, we compiled all the radial velocities for objects with
spectral types of M7 or later from \citet{2003ApJ...583..451M},
\citet{2004A&A...419..703B}, and \citet{2007ApJ...666.1205Z} and
computed space velocities for those objects with parallaxes and proper
motions from \citet{2002AJ....124.1170D}, \citet{1992AJ....103..638M},
\citet{2003AJ....126..975T}, \citet{2004AJ....127.2948V},
\citet{1996MNRAS.281..644T}, \citet{1995AJ....110.3014T}, and
\citet{2006AJ....132.2360H}. The resulting heliocentric velocities are
shown in Figure~\ref{fig:uvw}. The mean and rms scatter of the space
velocity of this population is $(U,V,W)$~=~($-$18$\pm$37,
$-$24$\pm$26, $-$5$\pm$15)~\kms, and \lhsint\ is only 1.3$\sigma$ away
from the mean of this ellipsoid. Thus, the space motion of \lhsint\ is
not significantly different from other ultracool dwarfs, implying an
age consistent with the population of ultracool dwarfs as a whole.
Several authors have attempted to estimate the age of this population,
typically comparing the distribution of tangential velocities ($V_{\rm
  tan}$, which requires only a proper motion and distance
determination) to the well studied nearby populations of FGKM stars.
The resulting age for the population of ultracool dwarfs estimated in
this way has been found to be 2--4~Gyr \citep{2002AJ....124.1170D,
  2009AJ....137....1F}.\footnote{\citet{2007ApJ...666.1205Z}
  determined a somewhat younger age ($\sim$1~Gyr) for the population
  of ultracool dwarfs, based on the small sample of L and T~dwarfs
  will full space velocities (21 objects). However, since L and
  T~dwarfs span a wider range of masses than earlier type objects, a
  typical IMF that rises at lower masses will naturally increase the
  number of young objects in this sample, biasing a kinematically
  derived age \citep[e.g., see Section~4.5
  of][]{2002AJ....124.1170D}.}

\begin{figure*}[t]
\plotone{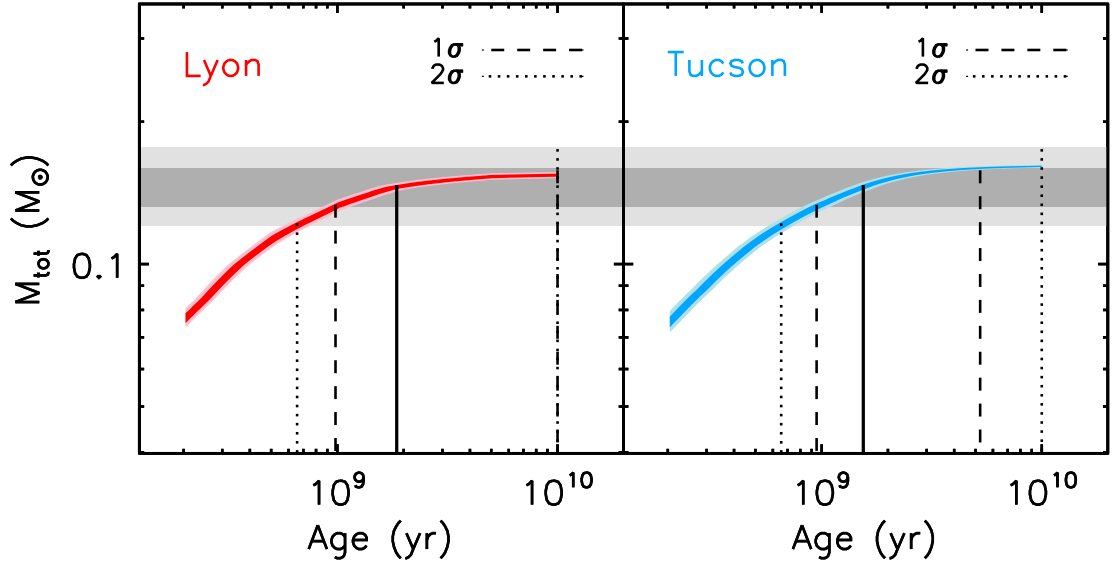}
\caption{ \normalsize Total mass (\Mtot) predicted by evolutionary
  models as a function of age, given the observational constraint of
  the luminosities of the two components of \lhsbin. The curved shaded
  regions show the 1$\sigma$ and 2$\sigma$ ranges in this
  model-derived mass. By applying the additional constraint of the
  measured total mass (\Mtot), we used the models to determine the age
  of \lhsbin\ (see Section~\ref{sec:modelage}). The horizontal gray
  bars show our 1$\sigma$ and 2$\sigma$ constraints on the total mass,
  and the resulting median, 1$\sigma$, and 2$\sigma$ model-inferred
  ages are shown by solid, dashed, and dotted lines, respectively.
  Model-inferred ages are truncated at 10~Gyr (the oldest age included
  in both sets of models) -- this happens at 1$\sigma$ for the Lyon
  models and 2$\sigma$ for the Tucson models.  This is because the
  high mass tail of the \Mtot\ distribution corresponds to both \lhsA\
  and \lhsB\ being stars at the bottom of the main sequence, and
  evolutionary models do not reach faint enough luminosities for such
  high mass objects by 10~Gyr. \label{fig:mtot-age}}
\end{figure*}

We have also assessed \lhsint's membership in the Galactic populations
of the thin disk \citep[1--10~Gyr; e.g.,][]{1998ApJ...497..870W} and
thick disk \citep[$\sim$10~Gyr; e.g.,][]{2002A&A...394..927I} using
the Besan\c{c}on model of the Galaxy \citep{2003A&A...409..523R}.
Because the scale height of these structures are much larger than the
measured distances to ultracool dwarfs, we found that a specialized
model population for the volume within 1 square-degree on the sky and
5~pc in the radial direction of \lhsint\ had essentially the same
space motions as the model population of all objects within 50~pc over
the whole sky. \citet{2003A&A...409..523R} describe the velocity
ellipsoids of the Besan\c{c}on model thin and thick disks (shown in
Figure~\ref{fig:uvw}) and how these space motions are allowed to
evolve over time due to dynamical heating. To determine membership
probabilities of a given object belonging to the thin and thick disks,
we simply counted the nearest 10$^3$ model objects in $(U, V,
W)$-space and computed the fraction that belonged to each population.
For \lhsint, we found a membership probability of 99.2\% for the thin
disk and 0.8\% for the thick disk.\footnote{We also computed
  membership probabilities for all the ultracool dwarfs shown in
  Figure~\ref{fig:uvw} and found only three systems with $>$~10\%
  probability of belonging to the thick disk: ESO~207-61 (69\%),
  GJ~1001 (40\%), and Gl~802 (36\%).}

Finally, the fact that \lhsint\ is an active H$\alpha$ flare star
\citep[$\log(L_{\rm H\alpha}/L_{\rm
  bol})$~=~$-$4.22;][]{2000AJ....120.1085G} could also potentially
provide an age constraint, as the activity of M~dwarfs changes with
age. \citet{2008AJ....135..785W} showed that the fraction of active
M~dwarfs as a function the vertical distance above the Galactic plane
($z$) provides a constraint on the activity lifetime of M~dwarfs,
given a model of how thick disk heating pumps up $z$ over time.
\citet{2008AJ....135..785W} found that the activity lifetime increases
monotonically with M~dwarf spectral type, and the latest type for
which they were able to determine a robust lifetime was M7
(8.0$^{+0.5}_{-1.0}$~Gyr). This provides a weak constraint on the age
of \lhsint, as its activity is therefore expected to last for at least
$\gtrsim$8~Gyr.

In summary, the space motion of \lhsint\ indicates that it is (1)
consistent with the field ultracool dwarf population and (2) most
likely (99.2\%) a member of the thin disk. These properties do not
strongly constrain its age: the age of the ultracool field dwarf
population is poorly constrained by current observations, and the thin
disk is believed to have been forming stars continuously over the past
$\sim$10~Gyr. The activity of \lhsint\ is expected to be long-lived,
and so it also does not strongly constrain the age. Thus, the
1$\sigma$ age range of 1.0--10~Gyr derived from evolutionary models
for \lhsint\ (Section~\ref{sec:modelage}) is consistent with the broad
range of ages allowed by its space motion and activity.


\section{Tests of Models \label{sec:tests}}

Mass is the primary input parameter for evolutionary models of
ultracool dwarfs.  Thus, the strongest tests of these models are made
possible by objects with a direct observational constraint on their
mass.  In the following analysis, we have used our derived total mass
of \lhsbin\ along with its other measured properties to test
theoretical models.  We drew the measured properties of \lhsbin\ from
appropriate random distributions, carefully accounting for the
covariance between different quantities (e.g., \Mtot\ and \Lbol\ are
correlated through the distance).  We have chosen to use \Lbol\ rather
than \Teff\ as the basis of our model comparisons because values of
\Teff\ in the literature are invariably tied to either evolutionary or
atmospheric theoretical models in some way due to the lack of direct
radius measurements for brown dwarfs.  By using \Lbol, which only
depends on direct measurements of SEDs and distances, we have avoided
circular comparisons.  This approach follows our previous work for
\twomassbin\ \citep{2008ApJ...689..436L} and \hdbin\
\citep{2009ApJ...692..729D}.

In the following, we consider two independent sets of evolutionary
models: the Tucson models \citep{1997ApJ...491..856B} and the Lyon
DUSTY models \citep{2000ApJ...542..464C}, which are appropriate for
both components of \lhsbin, as their photospheres are expected to be
in the effective temperature range over which appreciable amounts of
dust exist.

\subsection{Model-Inferred Age \label{sec:modelage}}

As described in detail by \citet{2008ApJ...689..436L} and
\citet{2009ApJ...692..729D}, the total mass of a binary along with its
individual component luminosities can be used to estimate the age of
the binary from evolutionary models. This age estimate can be
surprisingly precise when both components are likely to be substellar
since their luminosities depend very sensitively on age. For \lhsbin,
which only has one substellar component, we derive an age of
1.5$^{+4.1}_{-0.6}$~Gyr from the Tucson models and an age of
1.8$^{+8.2}_{-0.8}$~Gyr from the Lyon models
(Figure~\ref{fig:mtot-age}).

Note that there is an upper limit of 10~Gyr on the ages estimated from
evolutionary models as this is the oldest age included in both sets of
models. Our age estimates for \lhsbin\ hit this limit at 1$\sigma$ for
the Lyon models because the 1$\sigma$ upper limit on total mass
(0.160~\Msun) corresponds to both \lhsA\ and \lhsB\ being stars at the
bottom of the main sequence, and the Lyon models do not reach faint
enough luminosities for such high mass objects by 10~Gyr. This is also
the case for the Tucson models, though the limit is reached beyond the
1$\sigma$ upper limit in the total mass. Using models that extend to
older ages (e.g., the age of the universe) would not remedy this
problem because \lhsB's luminosity is much too low for a star even
though the total mass formally allows it to be one (see
Section~\ref{sec:qratio}).

\subsection{Individual Masses \label{sec:qratio}}

By measuring the relative orbit of \lhsbin, we have determined its
total mass to a precision of 10\% (dominated by the error in the
parallax).  One way to estimate individual masses is to use
evolutionary models in the same way they were used to infer the system
age.  By constraining the model-derived individual masses of \lhsA\
and \lhsB\ to add up to the observed total mass, while still matching
their observed luminosities, the Tucson models give masses of
0.0839$^{+0.0007}_{-0.0015}$~\Msun\ and
0.061$^{+0.014}_{-0.011}$~\Msun\ for \lhsA\ and \lhsB, respectively.
The Lyon models give masses of 0.0848$^{+0.0010}_{-0.0012}$~\Msun\ and
0.060$^{+0.008}_{-0.012}$~\Msun. We can also estimate the mass ratio
($q$~$\equiv$~$M_{\rm B}/M_{\rm A}$) in this fashion, and the Tucson
models give $q$~=~0.73$^{+0.16}_{-0.12}$, while the Lyon models give
$q$~=~0.71$^{+0.09}_{-0.14}$. This is among the lowest field ultracool
binary mass ratios \citep[e.g., see][]{2007prpl.conf..427B} and is
consistent with the original estimate of 0.76 by
\citet{2003ApJ...584..453F} based on their photometry and estimated
age range (2--12~Gyr).

\tabletypesize{\small}
\begin{deluxetable}{lccc}[t]
\tablewidth{0pt}
\tablecaption{Evolutionary Model-derived Properties \label{tbl:model}}
\tablehead{
\colhead{Property}    &
\colhead{Median}      &
\colhead{68.3\% c.l.} &
\colhead{95.4\% c.l.} }
\startdata
                           \multicolumn{4}{c}{\bf Tucson Models \citep{1997ApJ...491..856B}} \\
                           \cline{1-4}
                           \multicolumn{4}{c}{System} \\
                           \cline{1-4}
Age (Gyr)\tablenotemark{a}    & 1.5     & $-    0.6, 4.1    $ & $-    0.9, 8.5    $ \\
$q$ ($M_{\rm B}/M_{\rm A}$)   & 0.73    & $-   0.12, 0.16   $ & $-   0.19, 0.18   $ \\
$\Delta$\Teff\ (K)            & 1130    & $-     40, 30     $ & $-     50, 60     $ \\
                           \cline{1-4}
                           \multicolumn{4}{c}{Component A} \\
                           \cline{1-4}
$M_{\rm A}$ (\Msun)           & 0.0839  & $- 0.0015, 0.0007 $ & $-  0.005, 0.001  $ \\
$T_{\rm eff,A}$ (K)           & 2580    & $-     30, 30     $ & $-     50, 50     $ \\
$\log(g_{\rm A})$ (cgs)       & 5.381   & $-  0.014, 0.009  $ & $-   0.04, 0.02   $ \\
$R_{\rm A}$ (\Rsun)           & 0.0978  & $- 0.0012, 0.0012 $ & $-  0.002, 0.003  $ \\
Li$_{\rm A}$/Li$_0$           & 0.00    & $-   0.00, 0.00   $ & $-   0.00, 0.00   $ \\
                           \cline{1-4}
                           \multicolumn{4}{c}{Component B} \\
                           \cline{1-4}
$M_{\rm B}$ (\Msun)           & 0.061   & $-  0.011, 0.014  $ & $-  0.019, 0.015  $ \\
$T_{\rm eff,B}$ (K)           & 1450    & $-     40, 40     $ & $-     80, 70     $ \\
$\log(g_{\rm B})$ (cgs)       & 5.35    & $-   0.13, 0.14   $ & $-   0.25, 0.15   $ \\
$R_{\rm B}$ (\Rsun)           & 0.0868  & $-  0.005, 0.005  $ & $-  0.005, 0.009  $ \\
Li$_{\rm B}$/Li$_0$           & 0.5     & $-    0.5, 0.5    $ & $-    0.5, 0.5    $ \\
                           \cline{1-4}
                           \multicolumn{4}{c}{} \\
                           \multicolumn{4}{c}{\bf Lyon Models \citep[DUSTY;][]{2000ApJ...542..464C}} \\
                           \cline{1-4}
                           \multicolumn{4}{c}{System} \\
                           \cline{1-4}
Age (Gyr)\tablenotemark{a}    & 1.8     & $-    0.8, 8.2    $ & $-    1.1, 8.2    $ \\
$q$ ($M_{\rm B}/M_{\rm A}$)   & 0.71    & $-   0.14, 0.09   $ & $-   0.21, 0.10   $ \\
$\Delta$\Teff\ (K)            & 1040    & $-     40, 50     $ & $-     50, 80     $ \\
                           \cline{1-4}
                           \multicolumn{4}{c}{Component A} \\
                           \cline{1-4}
$M_{\rm A}$ (\Msun)           & 0.0848  & $- 0.0012, 0.0010 $ & $-  0.004, 0.002  $ \\
$T_{\rm eff,A}$ (K)           & 2470    & $-     30, 30     $ & $-     50, 60     $ \\
$\log(g_{\rm A})$ (cgs)       & 5.307   & $-  0.008, 0.007  $ & $-   0.03, 0.01   $ \\
$R_{\rm A}$ (\Rsun)           & 0.1073  & $- 0.0010, 0.0011 $ & $-  0.002, 0.002  $ \\
Li$_{\rm A}$/Li$_0$           & 0.00    & $-   0.00, 0.00   $ & $-   0.00, 0.00   $ \\
                           \cline{1-4}
                           \multicolumn{4}{c}{Component B} \\
                           \cline{1-4}
$M_{\rm B}$ (\Msun)           & 0.060   & $-  0.012, 0.008  $ & $-  0.020, 0.009  $ \\
$T_{\rm eff,B}$ (K)           & 1430    & $-     50, 40     $ & $-     90, 70     $ \\
$\log(g_{\rm B})$ (cgs)       & 5.31    & $-   0.15, 0.09   $ & $-   0.28, 0.09   $ \\
$R_{\rm B}$ (\Rsun)           & 0.0895  & $- 0.0041, 0.0064 $ & $-  0.004, 0.012  $ \\
Li$_{\rm B}$/Li$_0$           & 0.0     & $-    0.0, 0.9    $ & $-    0.0, 1.0    $ \\
\enddata
\tablenotetext{a}{Both sets of evolutionary models are only computed
  up to an age of 10~Gyr; therefore, this defines the upper limit on
  the model-derived ages.}
\end{deluxetable}

We can also estimate the individual masses of \lhsbin\ by subtracting
the mass of the stellar primary from the total mass. There are only
three stars of comparable spectral type and absolute magnitude to
\lhsA\ with dynamical mass estimates: LHS~1070B \citep[M8.5,
$M_V=19.24\pm0.07$~mag,
$M_K=10.42\pm0.04$~mag;][]{1994A&A...291L..47L, 2000A&A...353..691L},
LHS~1070C \citep[M9, $M_V=19.62\pm0.08$~mag,
$M_K=10.76\pm0.04$~mag;][]{1994A&A...291L..47L, 2000A&A...353..691L},
and GJ~1245C \citep[$M_V=18.55\pm0.05$~mag,
$M_K=9.99\pm0.04$~mag;][]{1993AJ....106..773H}. The spectroscopically
unclassified object GJ~1245C is the closest in absolute magnitude to
\lhsA\ and the only one of these with a direct dynamical mass
measurement \citep[0.074$\pm$0.013~\Msun;][]{1999ApJ...512..864H}.
LHS~1070BC only has a total mass determined from its relative orbit
\citep[0.157$\pm$0.009~\Msun;][]{2008A&A...484..429S}. If we simply
assume that the mass of \lhsA\ is the same as GJ~1245C, then the
resulting mass estimate for \lhsB\ is 0.072$\pm$0.019~\Msun. This is
somewhat larger than ($+$0.6$\sigma$) but consistent with the mass
estimates from evolutionary models (0.061$^{+0.014}_{-0.011}$~\Msun\
and 0.060$^{+0.008}_{-0.012}$~\Msun), which have somewhat smaller
uncertainties.  This approach is limited by the small number and low
precision of dynamical masses for stars at the bottom of the
main sequence, which will be improved by future mass measurements for
late-M dwarf binaries.

The mass of \lhsB\ is estimated to be substellar as it is below
0.070--0.092~\Msun, the range in the hydrogen-fusing minimum mass for
plausible values of the metallicity and helium fraction \citep{bur01}.
This is not surprising since theoretical models predict that stars at
the very bottom of the main sequence could not be as faint as \lhsB.
We can therefore estimate an upper limit on the total mass from the
hydrogen-fusing minimum mass ($\approx$0.075~\Msun\ for solar
metallicity) and the highest mass expected for \lhsA. There are few
mass measurements at the bottom of the main sequence, but we estimate
that a conservative upper limit on the mass of \lhsA\ is 0.095~\Msun.
Gl~866C has a mass of 0.0930$\pm$0.0008~\Msun\ and at
$M_V$~=~17.4$\pm$0.4~mag is more than a magnitude brighter than \lhsA;
no objects with masses $>$~0.100~\Msun\ come within 2~mag of \lhsA's
$M_V$.  These upper limits on the masses of the individual components
of \lhsbin\ imply \Mtot~$<$~0.170~\Msun, 1.7$\sigma$ larger than our
derived dynamical mass. Thus, we suggest that the high-mass tail of
the \Mtot\ distribution, while formally allowed by the data, is
practically excluded based on the properties of the individual
components.\footnote{If \lhsint\ were a higher order multiple, then
  its total mass would be more consistent with the high-mass tail of
  the \Mtot\ distribution.  However, the fact that the luminosities
  and colors of both components agree well with single field dwarfs
  makes unresolved multiplicity unlikely.}

\subsection{Temperatures and Surface Gravities \label{sec:teff-logg}}

Without radius measurements for \lhsA\ and \lhsB, we cannot directly
determine their effective temperatures or surface
gravities.\footnote{Since \lhsA\ is a chromospherically active star in
  a binary system, it may be possible to estimate its radius using the
  technique employed by \citet{2009ApJ...695..310B} who measured the
  rotation period of 2MASSW~J0746425$+$200032A from its chromospheric
  radio emission and combined this with its $v\sin(i)$ and orbital
  inclination ($i$) to derive its radius. This method assumes that the
  orbital and rotation axes are aligned. With a $v\sin(i)$ of
  20~\kms\ \citep{2003ApJ...583..451M}, \lhsA\ is expected to have a
  rotation period of about 240 minutes.} However, we have used
evolutionary models to estimate these properties in the same fashion
as our model-derived age and individual masses. 

The Lyon models give effective temperatures for \lhsA\ and \lhsB\ of
2470$\pm$30~K and 1430$\pm$40~K, while the Tucson models give
systematically hotter but formally consistent temperatures of
2580$\pm$30~K and 1450$\pm$40~K. Because of the nearly flat
mass--radius relation for such low mass objects, it is essentially the
uncertainties in the component luminosities that determine the
precision in the model-derived effective temperatures for \lhsA. Since
brown dwarfs cool over time, the precision in \lhsB's effective
temperature also depends somewhat on the uncertainty in the age, which
in turn depends on the mass error. While our uncertainties were
derived specifically for this system, approximate scaling relations
derived by \citet{2008ApJ...689..436L} can provide simple estimates of
the relative dependencies of the mass and luminosity errors on the
uncertainty in \Teff.

The Lyon models give surface gravities for \lhsA\ and \lhsB\ of
\logg~=~5.307$^{+0.007}_{-0.008}$ and 5.31$^{+0.09}_{-0.15}$ (cgs),
while the Tucson models give systematically higher gravities of
\logg~=~5.381$^{+0.009}_{-0.014}$ and 5.35$^{+0.14}_{-0.13}$ (cgs).
Since the radius is essentially constant with age, the precision in
model-inferred surface gravity is driven by the precision in the
measured total mass. The difference between the two sets of
model-inferred surface gravities, which are formally inconsistent for
\lhsA, arises from small differences ($\lesssim$~9\%) in the
model-predicted radii (Table~\ref{tbl:model}).

\subsubsection{Comparison to Field Dwarfs \label{sec:teff-field}}

The effective temperatures we derive from evolutionary models for
\lhsA\ and \lhsB\ can be compared to those which have been determined
for other objects of similar spectral type. Such estimates in the
literature utilize the nearly flat mass--radius relationship predicted
by theoretical models for brown dwarfs, adopting either a typical age
\citep[e.g., 3~Gyr;][]{gol04} or radius \citep[e.g.,
0.90$\pm$0.15~\Rsun;][]{2004AJ....127.2948V}.

For \lhsA\ (M8$\pm$0.5), field M7--M9 dwarfs with \Lbol\ measurements
have estimated effective temperatures ranging between 1850--2650~K
\citep{2001ApJ...548..908L}.  For \lhsB\ (L7$\pm$1), field L6--L8
dwarfs with \Lbol\ measurements have estimated effective temperatures
ranging between 1300--1700~K.  Both of these broad ranges of effective
temperature are consistent with our model-inferred effective
temperatures.  However, since both estimates are based on evolutionary
models, this only means that field L6--L8 dwarfs from previous studies
encompass objects of the same mass/age as \lhsB.

\citet{2007ApJ...667..527G} used their 24~\micron\ \Spitzer/MIPS
photometry to determine effective temperatures for nine M7.5--M8.5
dwarfs (including \lhsint) using the nearly model independent infrared
flux method. These temperatures range from 2400--2730~K (excluding
\lhsint). They derived an effective temperature of 2380~K for \lhsint,
but \lhsB\ likely affects this estimate since it is not a negligible
source of flux at 24~\micron: the atmospheric models of
\citet{2005astro.ph..9066B} predict that it would contribute 0.16~mJy
of the measured 0.99~mJy at 24~\micron. This is likely why the
$\Ks-[24]$ color of \lhsint\ (1.09$\pm$0.10~mag) is redder than any
other object of its spectral type (mean and rms of 0.80$\pm$0.08~mag),
and this could explain why the temperature estimate for \lhsint\ is
outside the range of the other eight M7.5--M8.5 dwarfs in the
\citet{2007ApJ...667..527G} sample. The infrared flux method
temperature range of the other M7.5--M8.5 dwarfs (2400--2730~K) is in
excellent agreement with the evolutionary model inferred effective
temperatures for \lhsA.

\begin{figure}[b]
\plotone{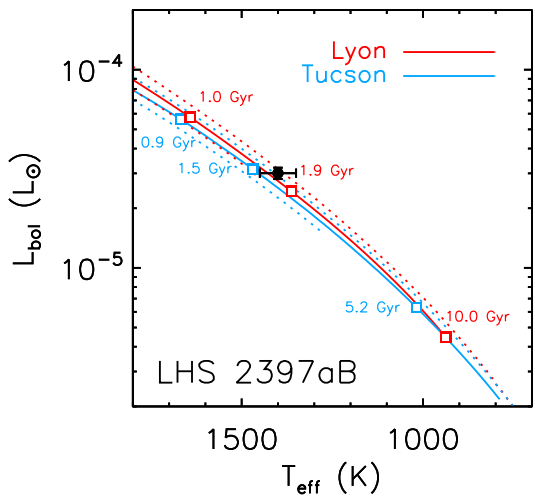}
\caption{\normalsize The Hertzsprung-Russell diagram showing isomass
  lines from evolutionary models for the mass of \lhsB\ with dotted
  lines encompassing the 1$\sigma$ uncertainty in the mass. The
  effective temperature of field L7$\pm$1 dwarfs as determined from
  spectral synthesis (1400~K) is shown as a filled circle with 50~K
  error bars, which correspond to the atmospheric model grid size. The
  open squares demarcate the median and 1$\sigma$ confidence limits on
  the evolutionary model-derived age of \lhsbin, using the combined
  constraint of the total mass and individual luminosities (see
  Figure~\ref{fig:mtot-age}). The atmospheric model temperature is in
  very good agreement with the evolutionary models, unlike previous
  studies which have found 100--200~K discrepancies on the H-R
  diagram \citep{2008ApJ...689..436L,
    2009ApJ...692..729D}. \label{fig:h-r}}
\end{figure}

\begin{figure*}
\plotone{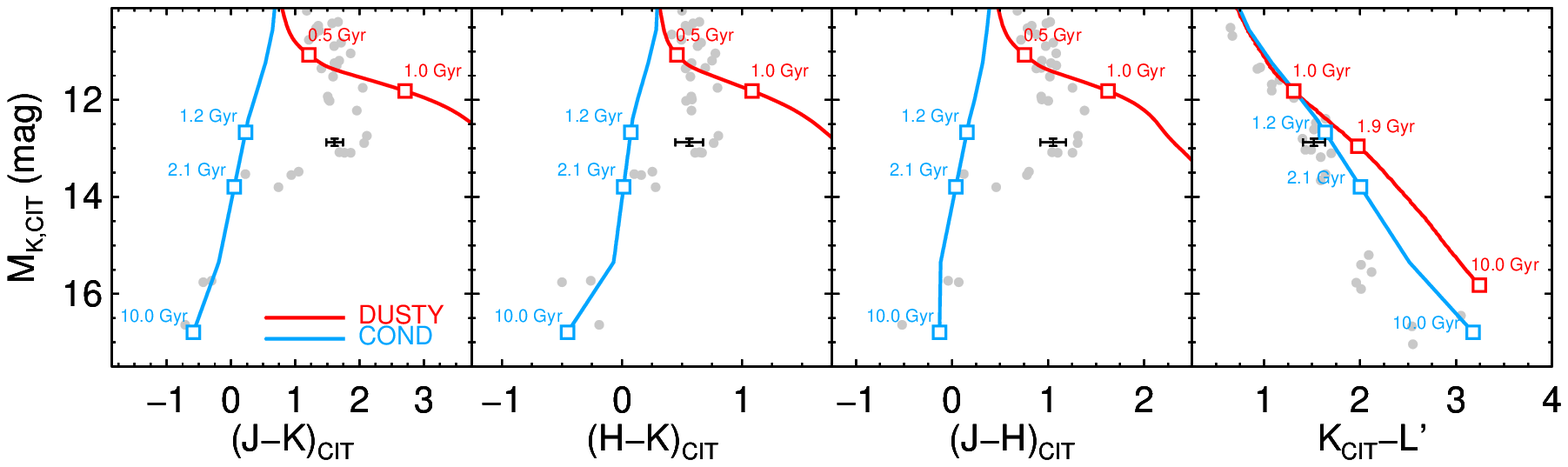}
\caption{ \normalsize Color-magnitude diagrams showing the measured
  properties of \lhsB\ compared to Lyon evolutionary tracks.  The
  solid lines are isomass tracks from the DUSTY
  \citep{2000ApJ...542..464C} and COND \citep{2003A&A...402..701B}
  models ($JHK$ photometry on the CIT system and \Lp-band photometry
  on the Johnson-Glass system).  The open squares demarcate ages along
  the isomass lines, showing the median and 1$\sigma$ confidence
  limits on the model-inferred age of \lhsbin\ if possible (0.5~Gyr
  also shown for reference).  Field dwarfs with parallax measurements
  better than 10\% and $JHK$ colors better than 0.10~mag are shown as
  filled gray circles.  \lhsB\ has near-infrared colors typical of
  field dwarfs, but neither set of evolutionary models reproduces the
  observed colors of \lhsB.  This is not surprising since each model
  is a limiting case in the amount of dust in the photosphere (very
  dusty or no dust) and \lhsB\ is expected to be intermediate given
  its estimated spectral type of L7$\pm$1. The $K-\Lp$ colors
  predicted by models are by far the least discrepant with the
  data. \label{fig:jhk-cmd}}
\end{figure*}

\subsubsection{Comparison to Atmospheric Models \label{sec:teff-atm}}

Effective temperatures have also been determined for many field dwarfs
through spectral synthesis modeling. By comparing these results to our
evolutionary model-derived temperatures, we perform a consistency
check between these two classes of models, which are essentially
independent. Even though atmospheric models provide an important
boundary condition for evolutionary models, very similar bulk
properties (e.g., \Lbol\ and \Teff) are predicted by evolutionary
models using widely varying boundary conditions
\citep{2000ARA&A..38..337C, 2008ApJ...689.1327S}.

Using spectral synthesis modeling over a very narrow spectral range
(2.297--2.310~\micron), \citet{2005MNRAS.358..105J} found effective
temperatures of 2900~K for the two M7--M9 dwarfs in their study. Using
a broader spectral range (1.0--2.5~\micron),
\citet{2001ApJ...548..908L} found much cooler temperatures of
2100--2300~K for the five M7--M9 dwarfs in their study. The
evolutionary model-inferred range of effective temperatures for \lhsA\
(2440--2610~K) is inconsistent with both results from spectral
synthesis modeling. Both authors cite the treatment of dust as the
likely cause of discrepancies they found between the atmospheric
models and their spectra.

Using spectral synthesis modeling over a very broad spectral range
(0.95--14.5~\micron), \citet{2008ApJ...678.1372C} derived effective
temperatures for several field L and T dwarfs, including two late-L
dwarfs of comparable spectral type and $JHK$ color to \lhsB. These two
objects (2MASS~J0825$+$2115, L7.5; DENIS~J0255$-$4700, L8) were both
found to have effective temperatures of 1400~K, which is in perfect
agreement with our model-derived temperature estimates for \lhsB. This
is somewhat surprising as previous studies of L and T~dwarfs have
found large (100--200~K) discrepancies between temperatures derived
from atmospheric and evolutionary models \citep{2008ApJ...689..436L,
  2009ApJ...692..729D}. Figure~\ref{fig:h-r} shows the atmospheric
model predicted temperature for \lhsB\ on the Hertzsprung-Russell
(H-R) diagram with the evolutionary model tracks. It is interesting to
note that both components of the L4+L4 binary \hdbin\ were found to
lie above the evolutionary model tracks \citep{2009ApJ...692..729D},
both components of the T5+T5.5 binary \twomassbin\ were found to lie
below the tracks \citep{2008ApJ...689..436L}, and \lhsB\ with an
intermediate spectral type of L7 falls exactly on the tracks. This
suggests that whatever different systematic errors are harbored in
each class of models may cancel out on the H-R diagram at this
spectral type.

\citet{2008ApJ...678.1372C} also derived surface gravities for the two
late-L dwarfs in their study by direct model fitting
(\logg~=~4.5--5.5) and by using their evolutionary sequences
(\logg~=~5.4--5.5). The latter values were determined from the
atmospheric model \Teff\ and the radius derived from the flux
normalization constant $(R/d)^2$, which yields the radius if the
distance is known. This is because atmospheric models predict the
emergent flux in absolute units, which can be compared directly to
flux-calibrated spectra. With independent estimates of both $R$ and
\Teff\ (from atmosphere fitting), any other property can be derived
using evolutionary models. The surface gravities derived by
\citet{2008ApJ...678.1372C} in this way for these two objects are
higher by 0.4--1.9$\sigma$ than our evolutionary model inferred values
for \lhsB\ from Lyon and Tucson models (\logg~=~5.31--5.34).

\subsection{Near-Infrared Colors \label{sec:cmd}}

The Lyon evolutionary models provide predictions of the fluxes of
ultracool dwarfs in standard filter bandpasses as a function of model
mass and age.\footnote{The models give $JHK$ photometry on the CIT
  system, and we convert our measured MKO system colors to this system
  using the relations of \citep{2004PASP..116....9S}. The model
  \Lp-band photometry is given on the Johnson-Glass system, and no
  empirical relations between this and the MKO photometric system
  exist, so we neglect any differences.} We have derived the
model-predicted near-infrared colors of both components of \lhsbin\ in
the same fashion as the individual masses, effective temperatures, and
age (i.e., using the combined observational constraints of the total
mass and individual luminosities). For \lhsA, the Lyon DUSTY models
predict a very narrow range of colors: $J-K = 0.784\pm0.007$~mag, $H-K
= 0.286\pm0.002$~mag, and $J-H = 0.498\pm0.006$~mag. None of these are
consistent with the observed colors of \lhsA: $J-K = 1.28\pm0.03$~mag,
$H-K = 0.43\pm0.03$~mag, and $J-H = 0.85\pm0.03$~mag. All of the
observed colors are redder than predicted, with $J-K$ being the most
discrepant (0.50~mag), followed by $J-H$ (0.35~mag) and $H-K$
(0.14~mag). Since the $J-K$ color of \lhsA\ is consistent with other
field dwarfs of spectral type M8 \citep[1.14$\pm$0.10~mag;
e.g.,][]{2009AJ....137....1F}, the Lyon DUSTY models will generally
provide inaccurate estimates of the fundamental properties of late-M
dwarfs from their near-infrared colors. In fact, it would not be
possible to estimate the mass of \lhsA\ from the color--magnitude
diagram as no model isochrone is consistent with the data.

We show the model-predicted $JHK\Lp$ colors of \lhsB\ compared to the
observations and other field dwarfs in Figure~\ref{fig:jhk-cmd}. The
colors of \lhsB\ are fully consistent with field dwarfs of similar
$K$-band absolute magnitude but are significantly bluer than the DUSTY
models for any assumed age. This discrepancy with the models is not
surprising as it is well known that the DUSTY models do not accurately
reproduce the near-infrared colors of late-L dwarfs. This is at least
partly because DUSTY models represent the limiting case of very high
dust content in the photosphere, while late-L dwarfs are actually in
transition to dust-free atmospheres. For comparison, we also show the
COND models \citep{2003A&A...402..701B}, which represent the converse
limiting case of a dust-free photosphere. As such, the COND models
reproduce the colors of the bluest T~dwarfs well, in which dust has
sedimented below the photosphere, but they fail to reproduce the
colors of \lhsB. For both DUSTY and COND models, the least discrepant
color is $K-\Lp$, which is only about 0.5~mag bluer than predicted by
DUSTY or COND, presumably because the effects of dust are less severe
at longer wavelengths. The difficulty in modeling the colors of
objects like \lhsB\ that are in transition from the L to T spectral
class is well known, though models developed recently that attempt to
account for the sedimentation of dust have enjoyed some success at
reproducing observations \citep{2005astro.ph..9066B,
  2008ApJ...689.1327S}.


\section{Discussion \label{sec:discuss}}

\subsection{Effective Temperature of the L/T Transition  \label{sec:teff-LT}}

Measurements of the fundamental properties of brown dwarfs undergoing
the transition from L to T dwarfs are critical for understanding the
physics governing dust in ultracool atmospheres.  With a spectral type
of L7$\pm$1, \lhsB\ is one such object for which our mass and
luminosity measurements have enabled precise effective temperature and
surface gravity estimates (Section~\ref{sec:teff-logg}).  In fact,
\citet{2008ApJ...689..436L} noted that ``mass benchmarks'' such as
\lhsB\ provide much more precise temperature estimates than ``age
benchmarks'' (i.e., substellar companions to stars of known age) since
the ages of field stars are notoriously difficult to estimate
precisely \citep[e.g., see][and references therein]{mam08-ages}.
\citet{2006ApJ...651.1166M} suggested that the effective temperature
of the L/T transition varies with surface gravity (or equivalently,
age), based on such age benchmark objects.  Thus, it is interesting to
compare the estimated temperature of \lhsB\ to those of the few late-L
and early-T dwarf companions with age estimates: Gl~584C
\citep[L8;][]{2001AJ....121.3235K}, HD~203030B
\citep[L7.5;][]{2006ApJ...651.1166M}, and HN~Peg~B
\citep[T2.5;][]{2006astro.ph..9464L}.\footnote{Gl~337CD
  \citep[L8+T;][]{wils01, 2005AJ....129.2849B} is another such
  companion; however, it is a tight binary with poor constraints on
  its resolved properties.  Thus, we exclude it from our discussion.}

For Gl~584C, \citet{gol04} derived an effective temperature range of
1300--1400~K from evolutionary models, using its measured luminosity
($\log{\Lbol/\Lsun}$~=~$-$4.58$\pm$0.04) and estimated age range
(1--2.5~Gyr).  This is consistent with our temperature estimate for
\lhsB\ (1380--1470~K from Lyon models; 1410--1490~K from Tucson
models).  In a similar fashion, \citet{2006ApJ...651.1166M} and
\citet{2006astro.ph..9464L} estimated the effective temperature ranges
for two companions to young stars: 1090--1280~K for HD~203030B
(130--400~Myr) and 1060--1200~K for HN~Peg~B (100--500~Myr).  Compared
to Gl~584C and \lhsB, the temperatures of these young companions are
much cooler.  As discussed by \citet{2006ApJ...651.1166M}, field
dwarfs of similar spectral type (L7--L9) with measured luminosities
also have higher estimated temperatures \citep[mean and rms of
1460$\pm$100~K; from][]{2004AJ....127.2948V}, though these estimates
depend strongly on the estimated age of the field population, which is
not well-constrained observationally.

\citet{2006ApJ...651.1166M} suggested two possible explanations for
the observed differences in effective temperatures: (1) the age of the
field population has been overestimated and (2) the effective
temperature of the L/T transition depends on surface gravity. If field
dwarfs are younger than estimated, their radii will be larger and thus
lower effective temperatures are required to match their measured
luminosities. Support for this possibility has come from
\citet{2008ApJ...689..436L}, who suggested that the age of field
dwarfs may be overestimated by as much as a factor of 6$\pm$3
(implying an age range of 0.3--1.0~Gyr for field objects) based on a
$\approx$100~K discrepancy between the field population and their
dynamical mass-based temperature estimates for the components of
\twomassbin\ (T5+T5.5). Such an overestimate would impact effective
temperature estimates at all spectral types, not just the L/T
transition. \citet{2006ApJ...651.1166M} preferred the explanation that
the L/T transition is gravity dependent, so that objects of different
radii/ages would undergo this evolutionary phase at different
effective temperatures. Our effective temperature determination for
\lhsB\ supports this idea since \lhsB\ belongs to the field population
(Section~\ref{sec:age}) and its temperature is inconsistent by
100--400~K with the young late-L companions HD~203030B and HN~Peg~B.

\subsection{Direct Measurement of the Mass Ratio  \label{sec:qdirect}}

Although the total dynamical mass we have measured enables strong
tests of theoretical models, these tests would be even more potent
with the direct measurement of the individual component masses. In
principle, this can be done using radial velocity or astrometric
monitoring, both of which provide a measurement of the binary's mass
ratio. The radial velocity semi-amplitude of \lhsA\ is 1.9~\kms\ for
the mass ratio of 0.7 derived from evolutionary models
(Section~\ref{sec:qratio}). Thus, a radial velocity precision of
0.1~\kms\ would be needed to achieve a mass ratio measurement precise
to 10\% (i.e., comparable to the precision in the total mass). As
discussed in Section~\ref{sec:orbit}, \citet{2006AJ....132..663B}
measured radial velocities for \lhsA\ at two epochs separated by
almost exactly half an orbital period, but with a precision of
1.6~\kms\ these data are not sufficient to determine the mass ratio.

To determine the individual masses from astrometric monitoring, the
orbit of \lhsA\ about the system's barycenter must be measured. The
size of this orbit is 89~mas for a mass ratio of 0.7. Therefore, an
astrometric precision of 5~mas in the optical, where \lhsB\ is
essentially invisible, is all that would be needed to measure the mass
ratio to $\approx$10\% ($\sigma_q/q = (1+q)\sigma_{a_1}/a_1$). Such a
measurement is quite feasible, though it requires long-term
observations given the orbital period of 14.2~years.

\subsection{Lithium Depletion \label{sec:lithium}}

The presence or absence of lithium has long been used as a tracer of
the internal structure of objects, providing constraints on mass
and/or age, because it is only destroyed by fusion at temperatures
above $\approx$2.5$\times$10$^6$~K. There is a mass limit below which
lithium is never destroyed in the fully convective interiors of very
low-mass stars, and this occurs at $\approx$0.06~\Msun\
\citep{1996ApJ...459L..91C}. The estimated mass of \lhsB\ is extremely
close to this limit, so 1$\sigma$ confidence limits on its
model-derived lithium abundance given in Table~\ref{tbl:model} range
from zero (fully depleted) to unity (no depletion). There are
currently no objects with direct mass estimates to provide
observational constraints on the lithium depletion boundary for brown
dwarfs. Thus, observations of lithium absorption at 6708~\AA\ for
\lhsB\ would provide a strong test of theoretical models of brown
dwarf interiors. In addition, such observations would provide an
independent constraint on the mass and age of \lhsB\ that would enable
a test of model-predicted luminosity evolution.  Lithium observations
are only possible with a repaired Space Telescope Imaging Spectrograph
(STIS) onboard \HST, and the next few years are an ideal time for such
a measurement as the orbital separation is currently approaching its
maximum (Figure~\ref{fig:orbit-sep-pa}).


\section{Conclusions}

We have determined the orbit of the M8+L7 binary \lhsbin\ using
relative astrometry spanning 11.8~years of its 14.2-year orbit. The
astrometry and corresponding errors used to derive this orbit were
thoroughly examined through Monte Carlo simulations. The resulting
best-fit orbit has a reduced $\chi^2$ of 1.04 and total mass of
0.146$^{+0.015}_{-0.013}$~\Msun. The error in the dynamical mass
(10\%) is dominated by the 3.0\% error in the parallax, which
translates into a 9.0\% error in the mass. The total mass is
consistent with the primary component being a star and the companion
being a brown dwarf, as expected from their spectral types. The
combined observational constraints of the total mass and individual
luminosities break the mass--age--luminosity degeneracy for the
substellar companion, enabling specific predictions of its properties
from evolutionary models. Tucson models \citep{1997ApJ...491..856B}
predict an age for the system of 1.5$^{+4.1}_{-0.6}$~Gyr, and the Lyon
\citep[DUSTY;][]{2000ApJ...542..464C} model-derived age of
1.8$^{+8.2}_{-0.8}$~Gyr is consistent. These ages are consistent with
the (weak) observational constraints on the system: \lhsint's space
motion implies it is a member of the thin disk \citep[1--10~Gyr;
e.g.,][]{1998ApJ...497..870W}, and although it is chromospherically
active, its expected activity lifetime is $\gtrsim$~8~Gyr
\citep{2008AJ....135..785W}.

LHS~2397aB (L7$\pm$1) is now the first mass benchmark at the
transition from L to T spectral types. This enables a precise estimate
of the effective temperature of this phase of substellar evolution
from theoretical models, which give 1450$\pm$40~K (Tucson) and
1430$\pm$40~K (Lyon). This is $\approx$200~K higher than temperature
estimates for late-L companions to young stars: HD~203030B
\citep[130--400~Myr;][]{2006ApJ...651.1166M} and HN~Peg~B
\citep[100--500~Myr;][]{2006astro.ph..9464L}. However, the effective
temperature of \lhsB\ is consistent with the 1300--1400~K range
estimated for the older late-L companion Gl~584C
\citep[1--2.5~Gyr;][]{2001AJ....121.3235K}. This supports the idea
originally proposed by \citet{2006ApJ...651.1166M} that the L/T
transition can occur over a wide range of effective temperatures
due to a surface gravity dependence on the process. In addition, the
temperature for \lhsB\ is consistent with estimates for field late-L
dwarfs (1440$\pm$100~K), and thus an overestimate of the age of the
field population (or underestimate of radii) is not required to
explain the range of temperatures observed at the L/T transition, as
previously suggested \citep{2006ApJ...651.1166M,
  2008ApJ...689..436L}.

We have determined the spectral type of \lhsB\ from its $JHK\Lp$
absolute magnitudes, using a novel technique that enables a
quantitative assessment of the spectral type uncertainty by accounting
for both measurement errors and the rms scatter in the empirical
relations. Comparing to other objects of similar spectral type and
color for which \citet{2008ApJ...678.1372C} have conducted spectral
synthesis analysis allows us to test atmospheric models for
consistency with evolutionary models. The \citet{2008ApJ...678.1372C}
study included two late-L dwarfs, and their derived effective
temperatures were both 1400~K, which is consistent with the effective
temperature of \lhsB\ we derived. This is somewhat surprising as
similar consistency tests by previous studies of binary brown dwarfs
have found large (100--200~K) discrepancies between effective
temperatures derived from atmospheric and evolutionary
models. However, it is interesting to note that \lhsB\ has a spectral
type intermediate between \hdbin\ for which the effective temperatures
are \emph{under}estimated by atmospheric models
\citep[L4+L4;][]{2009ApJ...692..729D} and \twomassbin\ for which the
temperatures are \emph{over}estimated
\citep[T5+T5.5;][]{2008ApJ...689..436L}. This suggests that \lhsB\ may
be at a temperature where atmospheric and evolutionary models are in
agreement, because the significant systematic errors in the models
cancel out.

Because of its extreme constrast ratio ($\gtrsim$~100:1 in the
optical), \lhsbin\ is one of the few ultracool binaries whose mass
ratio can be readily measured through astrometric monitoring. Given
the estimated mass ratio of 0.7, the barycentric orbit of \lhsA\ is
89~mas, which is easily measureable with current instrumentation. It
is unlikely that \lhsA's 14-year orbital motion significantly impacted
the original optical parallax measurement, except by introducing
additional scatter in the measurements. Thus, new astrometric
monitoring observations should improve the parallax precision, which
is essential for further testing of models. The total mass is
currently constrained to a precision of 10\%, dominated by the error
in the parallax, and at this precision the secondary component is
formally allowed (at 1--2$\sigma$) to be a star at the bottom of the
main sequence. A more precise mass measurement for this system would
enable much stronger constraints on all model-predicted properties,
including the amount of lithium depletion in \lhsB. Its model-derived
individual mass ($\approx$0.06$\pm$0.01~\Msun) is very close to the
theoretical mass-limit for lithium burning, which has never been
tested with direct mass measurements. A lithium measurement for a
brown dwarf of known mass would enable a unique constraint on
substellar models as lithium depletion proceeds rapidly and
independently of luminosity evolution. Such a resolved measurement of
the the lithium doublet at 6708~\AA\ for \lhsB\ is only possible with
\HST/STIS.

Mass benchmarks enable some of the strongest tests of theoretical
models by providing measurements of one of the most fundamental
physical parameters. We have determined the mass of an object in the
L/T transition for the first time -- a key step in understanding this
complex phase of substellar evolution. Our ongoing Keck LGS AO orbital
monitoring of L/T binaries, in which the spectral type of one
component is late-L and the other is early-T, will soon yield
dynamical masses for a variety of L/T transition objects. These will
provide mass benchmarks in the L/T transition for a broad range of
surface gravities, enabling even stronger tests of the evolution of
ultracool atmospheres.


\acknowledgments

We gratefully acknowledge the Keck AO team for their exceptional
efforts in bringing the AO system to fruition. It is a pleasure to
thank Antonin Bouchez, David LeMignant, Marcos van Dam, Randy
Campbell, Al Conrad, Jim Lyke, Hien Tran, Jason McIlroy, and Gary
Punawai and the Keck Observatory staff for assistance with the
observations. We are very thankful for the contribution of Peter
Tuthill in establishing aperture masking at Keck. We are grateful to
Brian Cameron for making available his NIRC2 distortion solution,
C\'{e}line Reyl\'{e} for customized Besan\c{c}on Galaxy models, and
Adam Burrows and Isabelle Baraffe for providing finely gridded
evolutionary models.  We have benefitted from discussions with Michael
Cushing about theoretical models and Thierry Forveille about space
motions and orbit fitting using \orbit.  We are indebted to Katelyn
Allers for assistance in obtaining and reducing IRTF/SpeX data.  We
also acknowledge the referee's helpful comments on the organization of
this paper.
Our research has employed the 2MASS data products; NASA's
Astrophysical Data System; the SIMBAD database operated at CDS,
Strasbourg, France; the SpeX Prism Spectral Libraries, maintained by
Adam Burgasser at \texttt{http://www.browndwarfs.org/spexprism}; and
the M, L, and T~dwarf compendium housed at \texttt{DwarfArchives.org}
and maintained by Chris Gelino, Davy Kirkpatrick, and Adam Burgasser
\citep{2003IAUS..211..189K, 2004AAS...205.1113G}.
TJD and MCL acknowledge support for this work from NSF grant
AST-0507833, and MCL acknowledges support from an Alfred P. Sloan
Research Fellowship.
Finally, the authors wish to recognize and acknowledge the very
significant cultural role and reverence that the summit of Mauna Kea has
always had within the indigenous Hawaiian community.  We are most
fortunate to have the opportunity to conduct observations from this
mountain.

{\it Facilities:} \facility{Keck II Telescope (LGS AO, NIRC2)},
\facility{\HST\ (WFPC2)}, \facility{VLT (NACO)},
\facility{Gemini-North Telescope (Hokupa`a), \facility{IRTF (SpeX)}}


\end{document}